\pdfoutput=1

\documentclass[11pt]{article}
\usepackage{jheppub}

\usepackage{amsmath}
\usepackage{amssymb}
\usepackage{graphicx,color,slashed,braket}
\usepackage{bbm} 
\usepackage{ifpdf}
\usepackage{comment}

\newcommand{\be}{\begin{equation}}
\newcommand{\ee}{\end{equation}}
\newcommand{\ba}{\begin{eqnarray}}
\newcommand{\ea}{\end{eqnarray}}
\newcommand{\nn}{\nonumber}
\newcommand{\lp}{\left(}
\newcommand{\rp}{\right)}
\newcommand{\ls}{\left[}
\newcommand{\rs}{\right]}

\newcommand{\e}{\textrm{e}}

\newcommand{\w}{\wedge}

\newcommand{\N}{\mathcal{N}}

\def\rmi{{\rm i}}
\def\K{{K\"{a}hler} }
\def\ib{{\bar \imath}}
\def\jb{{\bar \jmath}}
\newcommand{\Db}{\overline{D3}}
\newcommand{\A}{\mathcal{A}}
\renewcommand{\L}{\mathcal{L}}

\allowdisplaybreaks

\title{The supersymmetric anti-D3-brane action in KKLT}

\author[a]{Niccol\`o Cribiori,}
\author[a]{Christoph Roupec,}
\author[a]{Timm Wrase}
\author[b,c]{and Yusuke Yamada}

\affiliation[a]{Institute for Theoretical Physics, TU Wien,\\
Wiedner Hauptstrasse 8-10/136, A-1040 Vienna, Austria}
\affiliation[b]{Department of Physics, Graduate School of Science,
The University of Tokyo,\\ Hongo 7-3-1
Bunkyo-ku, Tokyo 113-0033, Japan}
\affiliation[c]{Research Center for the Early Universe (RESCEU), Graduate School of Science,\\ The University of Tokyo, Hongo 7-3-1
Bunkyo-ku, Tokyo 113-0033, Japan}

\emailAdd{niccolo.cribiori@tuwien.ac.at}
\emailAdd{christoph.roupec@tuwien.ac.at}
\emailAdd{timm.wrase@tuwien.ac.at}
\emailAdd{yamada@resceu.s.u-tokyo.ac.jp}

\preprint{RESCEU-5/19}
\notoc

\abstract{

\noindent
An anti-D3-brane plays a crucial role in the construction of semi-realistic cosmological models in string theory. Part of its action provides an uplift term that has been used to lift AdS solutions to phenomenologically viable dS vacua in the KKLT and LVS setups. In the last few years it has been shown that this uplift breaks supersymmetry spontaneously and can be described in the 4d $\N=1$ supergravity language by using constrained supermultiplets. Here we derive the complete 4d $\N=1$ supergravity action for an anti-D3-brane coupled to all closed string background fields. In particular we include the vector field, the scalar fields and all fermions that live on the anti-D3-brane. 
}

\begin{document}

\maketitle

\newpage

\section{Introduction}
The KKLT scenario~\cite{Kachru:2003aw} provides the first construction of dS vacua in string theory. The very existence of such solutions in quantum gravity has recently been questioned, see~\cite{Brennan:2017rbf,Danielsson:2018ztv, Palti:2019pca} for review articles and references. However, there is so far no generally agreed on flaw in the KKLT scenario and some past criticisms have already been refuted. Given this status, it is important to improve our understanding of the KKLT setup further. One such line of research has focused on the description of the anti-D3-brane that provides the uplift from a supersymmetric AdS vacuum to a dS vacuum. In particular, it has become apparent in the last few years that one can describe the anti-D3-brane in terms of a 4d $\N=1$ supergravity action. In this paper, we continue this endeavor by deriving the complete 4d $\N=1$ supergravity effective action for the KKLT scenario, including the anti-D3-brane and all of its world volume fields.

Our supersymmetric low energy effective action shows that supersymmetry in the KKLT setup is spontaneously broken. While this might have been expected because the anti-D3-branes used as uplift in the KKLT scenario are an excited state in a supersymmetric theory~\cite{Kachru:2002gs}, it was not until 2014 that it was understood how to write down a supergravity action that reproduces the anti-D3-brane uplift term \cite{Ferrara:2014kva}. The connection of this uplift term to the anti-D3-brane in the KKLT setup was then clarified in~\cite{Kallosh:2014wsa, Bergshoeff:2015jxa, Kallosh:2015nia,Garcia-Etxebarria:2015lif}. 

The subject of brane supersymmetry breaking started with \cite{Sugimoto:1999tx, Antoniadis:1999xk, Angelantonj:1999jh, Aldazabal:1999jr, Angelantonj:1999ms} and the connection to non-linear supersymmetry was first studied in \cite{Dudas:2000nv, Pradisi:2001yv}. All of these developments have broadened into a variety of different research directions and led to many interesting related results during the last few years, see for example~\cite{Bertolini:2015hua,Bandos:2015xnf,Dasgupta:2016prs,Vercnocke:2016fbt,Kallosh:2016aep,Bandos:2016xyu,Aoki:2016tod,Aalsma:2017ulu,Kallosh:2017wnt,GarciadelMoral:2017vnz,Cribiori:2017laj, Kitazawa:2018zys, Krishnan:2018udc,Aalsma:2018pll,Cribiori:2018dlc}. However, so far nobody has succeeded in writing down the full four-dimensional low energy effective supergravity action that includes all anti-D3-brane world volume fields in the KKLT background. This action consists of a bosonic part, containing the three complex world volume scalars and the U$(1)$ gauge field, and of a fermionic part containing the four 4d fermions. For flux compactifications this fermionic world volume action is currently only known to quadratic order in the fermions~\cite{Grana:2002tu, Grana:2003ek, Marolf:2003ye, Tripathy:2005hv, Martucci:2005rb, Bergshoeff:2013pia}.

While the bosonic action seems at first to be the easier part, it is actually the fermionic action that has been mostly studied during the last few years~\cite{Kallosh:2014wsa, Bergshoeff:2015jxa, Kallosh:2015nia,Garcia-Etxebarria:2015lif,Dasgupta:2016prs,GarciadelMoral:2017vnz}. In particular, it has been shown that one can do an orientifold projection that removes all of the bosonic degrees of freedom from the anti-D3-brane. The fermions together with the bosonic uplift term can then be combined into a Volkov-Akulov type action~\cite{Volkov:1973ix} and be described in the 4d $\N=1$ supergravity action via constrained chiral multiplets. Recently, the complete action for the GKP background fields and the four anti-D3-brane world volume fermions has been derived in~\cite{GarciadelMoral:2017vnz}. Here we extend this work by studying the full KKLT background and by including also the world volume scalar fields and the U$(1)$ gauge vector. Thus, we derive the complete low energy effective supergravity action for an anti-D3-brane in the KKLT background.

The organization of the paper is as follows. In section~\ref{sec:antibraneaction} we review the action for an anti-D3-brane in the GKP and KKLT background. In section~\ref{sec:sugracomponent} we discuss the constrained multiplets in 4d $\N=1$ supergravity that we need to describe the anti-D3-brane. In section~\ref{sec:sugraaction} we derive the four dimensional $\N=1$ supergravity action for an anti-D3-brane in the KKLT background. We summarize our findings in section \ref{sec:summary} and we draw the conclusions in section~\ref{sec:conclusion}. Two appendices provide technical details.

\section{The anti-D3-brane action in the GKP and KKLT background}\label{sec:antibraneaction}
In this section we will review and (re-)derive the action for an anti-D3-brane in the GKP~\cite{Giddings:2001yu} and KKLT~\cite{Kachru:2003aw} backgrounds. While many aspects of this action have been studied before, we will include here all world volume fields of the anti-D3-brane and their couplings to the background moduli fields, which are the axio-dilaton $\tau = C_0 + \rmi e^{-\phi}$, the single \K modulus $T$ and the complex structure moduli $U^A$. The Dp-brane action in flux backgrounds is only understood up to quadratic order in the fermions~\cite{Grana:2002tu, Grana:2003ek, Marolf:2003ye, Tripathy:2005hv, Martucci:2005rb, Bergshoeff:2013pia}. The known pieces of the action therefore include a bosonic part and a fermionic part that is quadratic in the worldvolume fermions.\footnote{This action does not include terms that are linear in a worldvolume fermion and a closed string fermion.} We will discuss these separately in the following two subsections.

\subsection{The bosonic action}\label{sec:bosonic}
The bosonic anti-D3-brane action is the sum of the DBI-action and the Chern-Simons action and is given in 4d Einstein frame by
\ba
S^{\Db}_{bos} &=& S^{\rm DBI} + S^{\rm CS}\,,\\
S^{\rm DBI} &=& - \int d^4x \sqrt{-\det \lp {\rm P}\ls g_{\mu\nu} + e^{-\frac{\phi}{2}} B_{\mu\nu}\rs + e^{-\frac{\phi}{2}} F_{\mu\nu} \rp}\,,\\
S^{\rm CS} &=& -\int {\rm P} \ls (C_0 + C_2 + C_4) \w e^{B_2} \rs \w e^{F}\,.
\ea
Here we have set $l_s=2\pi\sqrt{\alpha'}=1$, while $B_2$ denotes the NSNS Kalb-Ramond field, $F_{\mu\nu }$ the field strength of the U$(1)$ gauge field living on the brane and P is the pullback to the brane world volume. To simplify the presentation, we have rescaled the U$(1)$ field strength by $2\pi$ with respect to the textbook by Polchinski \cite{Polchinski:1998rr}, i.e. $F_{\mu\nu }^{\rm Polchinski} = 2\pi F_{\mu\nu }^{\rm us}$. We have also rescaled the action by $1/2\pi$ to remove the brane tension $T_3=(2\pi)^{-3} (\alpha')^{-2}=2\pi$.

In a GKP background~\cite{Giddings:2001yu} the metric is warped and the presence of warping makes the identification of the \K modulus, Im$(T)$ in our case, rather cumbersome~\cite{Frey:2008xw}. For a single \K modulus there is a fixed overall scaling with respect to the volume for all of the terms in the action. We can identify this scaling by working with the following metric in Einstein frame~\cite{Giddings:2005ff, deAlwis:2016cty}
\be
\label{gdeAlwis}
ds^2 = e^{-6u(x)}\left(1+\frac{e^{-4\mathcal{A}(z)}}{e^{4u(x)}}\right)^{-\frac12} g_{\mu\nu} dx^\mu dx^\nu + e^{2u(x)}\left(1+\frac{e^{-4\mathcal{A}(z)}}{e^{4u(x)}}\right)^{\frac16} g_{a\bar b} dz^a dz^{\bar b}\,,
\ee
where the external indices are labeled by $\mu,\nu=0,1,2,3$, the internal indices by $a, \bar b=1,2,3$ and $e^{6u}=vol_6$ is the volume of the internal manifold, whose dependence on Im$(T)$ is going to be specified below.\footnote{The Ansatz \eqref{gdeAlwis} does not solve the mixed components of the 10d Einstein equations with one internal and one external index and one has to introduce a compensator field~\cite{Frey:2008xw}. This subtlety will not affect our result. However, it would be important to confirm this explicitly by doing a proper dimensional reduction of the anti-D3-brane extending the result for a supersymmetric D3-brane of \cite{Cownden:2016hpf}.}
This metric interpolates between the unwarped bulk region and the warped throat. We will be interested in the strong warping regime, namely $e^{-4\mathcal{A}}\gg e^{4u}$, where the metric reduces to
\be
ds^2 = e^{2\A(z)-4u(x)}g_{\mu\nu} dx^\mu dx^\nu + e^{\frac43u(x)-\frac23 \A(z)}g_{a\bar b} dz^a dz^{\bar b}.
\ee

We can now proceed and start to evaluate the DBI action. Following~\cite{McGuirk:2012sb}, this becomes\footnote{This action can be recast in the conventions used in~\cite{GarciadelMoral:2017vnz} by sending $\A \to \A - u$ and then $g_{a \bar b}\to g_{a \bar b}e^{-\frac43\A}$.}
\be\label{eq:DBIaction}
\begin{aligned}
S^{\rm DBI} = -  \int d^4x \sqrt{-g_4}\bigg( &e^{4 \A(H,\bar H)-8 u(x)} + \frac12 e^{\frac43\A(H,\bar H)-\frac83u(x)} g_{a\bar b}(H, \bar H) \partial_\mu H^a  \partial^\mu \bar H^{\bar b}\\
& +\frac{e^{-\phi(H, \bar H)}}{4} F_{\mu\nu} F^{\mu\nu} +\ldots \bigg)\,,
\end{aligned}
\ee
where the dots denote higher order terms. These are corrections, which are small with respect to the couplings that we wrote down explicitly. The warp factor, the internal metric and the dilaton are functions of the world volume scalars $H^a$, that indicate the position of the brane and that enter the action via the pullback P. We will assume that the brane sits at some position in the strongly warped region, but we will not need to specify it further. For the rest of our discussion we will consider small fluctuations around such a position and we indicate them with the same letter $H^a$ for convenience.

The kinetic term for the scalar fields arises entirely from the DBI part of the action and is therefore the same for D3-branes and anti-D3-branes. The rewriting of this term in 4d $\N=1$ supergravity was first discussed in~\cite{DeWolfe:2002nn}. There it was argued that such a kinetic term stems from a \K potential of the type
\be\label{eq:Kpot}
K = -3 \log\ls-\rmi (T-\bar{T}) + k(H,\bar H) \rs\,, 
\ee
where $T$ is our single \K modulus and $k(H,\bar H)$ is the \K potential corresponding to the internal Calabi-Yau metric $\partial_{H^a}\partial_{\bar{H}^{\bar b}} k(H,\bar H) \approx \frac{1}{6} e^{\frac43 (\A+u)}g_{a\bar b}$, where we neglected subleading terms (cf. appendix B of \cite{Baumann:2007ah}). The \K potential $k(H,\bar H)$ does not break the no-scale structure and enters the expression of the overall volume, which indeed depends on the open and closed string moduli via 
\be
vol_6 = e^{6u}= \lp -\rmi (T-\bar{T})+k(H,\bar H)\rp^{\frac32}.
\ee

The DBI action gives also rise to a scalar potential and a standard Maxwell term for the U$(1)$ gauge field, with coupling constant determined by Im$(\tau)=e^{-\phi}$ evaluated at the position of the brane. We will discuss both of these terms further when we combine them with the CS-action.

We now look at the CS-action for the anti-D3-brane. In the GKP background it reduces to
\ba\label{eq:CSaction}
S^{\rm CS} &=& - \int \left(\frac12 C_0(H,\bar H)  F \w F + C_4 (H, \bar H)\right)\nonumber\\
&=& - \int\left(\frac12 C_0(0,0) F \w F + C_4 (0,0) +\ldots\right)\\
&=& - \int d^4x \sqrt{-g_4}\left(-\frac{\text{Re}(\tau)}{8} \frac{\epsilon^{\mu\nu\rho\sigma}}{\sqrt{-g_4}}F_{\mu\nu}F_{\rho\sigma} + \alpha(H,\bar H)+\dots\right)\nonumber,
\ea
where in the second line we expanded around the position of the brane, $H^a=0$, and omitted higher order terms. We are using the fact that Re$(\tau)=C_0(0,0)$ and $C_4 = \alpha(z,\bar z) \sqrt{-g_4}dx^0\w dx^1 \w dx^2 \w dx^3$, where $g_4$ is the determinant of the unwarped four-dimensional metric. Recall that $C_2$ and $B_2$ with indices along the non-compact spacetime directions are projected out by the orientifold projection.

A D3-brane in the background we are considering preserves linear $\N=1$ supersymmetry in 4d. The U$(1)$ gauge field on its world volume has a gauge kinetic function given by $f(\tau) =- \rmi \, \tau$. This function is and has to be holomorphic and depends only on the axio-dilaton modulus $\tau$. The real part, Re$(f(\tau)) = {\rm Im}(\tau)=e^{-\phi}$, controls the coupling in the Maxwell term and the imaginary part, Im$(f(\tau)) =- {\rm Re}(\tau)=-C_0$, controls the theta term. This leads to an immediate problem in the case of the anti-D3-brane. With respect to the D3-brane the anti-D3-brane has a sign difference in the CS-action and thus, in order to get the correct sign for the theta term in formula~\eqref{eq:CSaction}, it seems that we would have to make the gauge kinetic function anti-holomorphic, $f(\bar \tau) = \rmi\, \bar \tau$. This however would not be compatible with supersymmetry, since the closed string field $\tau$ is part of an unconstrained chiral multiplet. 
We will show how to resolve this puzzle and maintain a holomorphic gauge kinetic function in section \ref{sec:sugraaction}. The crucial point will be that, in the background we are considering, an anti-D3-brane preserves \emph{non-linear} $\N=1$ supersymmetry in 4d.

We can now put together the two pieces and obtain the bosonic action of the anti-D3-brane. 
The second term in the CS-action \eqref{eq:CSaction} combines with the first term in the DBI-action~\eqref{eq:DBIaction} into what is usually denoted by
\be
\Phi_\pm \equiv e^{4 \A(H,\bar H)-8 u} \pm \alpha(H,\bar H),
\ee
with the plus being for the anti-D3-brane and the minus for the D3-brane. The equations of motion in the GKP solution enforce $\Phi_-=0$, so that the potential for a D3-brane, $V_{D3}= \Phi_-$, vanishes because the DBI-part and the CS-part exactly cancel. For an anti-D3-brane the contributions simply add up and we have
\begin{equation}\label{eq:bosonic}
\begin{aligned}
S^{\Db}_{bos} &= S^{DBI} + S^{CS}\\
&=- \int d^4x \sqrt{-g_4}\bigg(2  e^{4 \A(H,\bar H)-8 u}+\frac12 e^{\frac43\A(H,\bar H)-\frac83u(x)}g_{a\bar b}\partial_\mu H^a \partial^\mu \bar H^{\bar b} \\
&\qquad\qquad\qquad\qquad+ \frac{{\rm Im}(\tau)}{4}F_{\mu\nu}F^{\mu\nu} - \frac{{\rm Re}(\tau)}{8}\frac{\epsilon^{\mu\nu\rho\sigma}}{\sqrt{-g_4}}F_{\mu\nu}F_{\rho\sigma} +\dots\bigg),
\end{aligned}
\end{equation}
where we can identify the scalar potential
\be
\label{eq:scalarpot}
V_{\Db}(H, \bar H) =  \Phi_+=2  e^{4 \A(H,\bar H)-8 u}.
\ee
Since the warp factor is minimized at the bottom of the warped throat, this is likewise the point where also the anti-D3-brane potential is minimized and where the brane is dynamically attracted to. We can then expand the scalar potential around this point to obtain
\be
\begin{aligned}
V_{\Db}(H,\bar H) = 2 e^{4 \A_0-8 u_0}(&1 + 4 H^a \bar{H}^{\bar b} \partial_{H^a} \partial_{\bar{H}^{\bar b}} \A|_{H=0}+ 2 H^a H^b \partial_{H^a} \partial_{H^b} \A|_{H=0}\\
&+2 \bar{H}^{\bar a} \bar{H}^{\bar b} \partial_{\bar{H}^{\bar a}} \partial_{\bar{H}^{\bar b}} \A|_{H=0} +\ldots),
\end{aligned}
\ee
where $\A_0 \equiv \A|_{H=0}$, $u_0 \equiv u|_{H=0}$ and the dots stand for higher order terms, which are actually suppressed by the string scale (following the same logic as in section 3.1 of \cite{McGuirk:2012sb}). The first contribution in the expansion is the uplift term for an anti-D3-brane which, in a highly warped region, scales like $1/(vol_6)^\frac43$  \cite{Kachru:2003sx}.\footnote{In the unwarped region the uplift term scales actually like $1/(vol_6)^2$~\cite{Kachru:2003aw}.}
In section~\ref{sec:sugraaction} we will show how to reproduce the above scalar potential using a modified version of the \K potential given in~\eqref{eq:Kpot}.

In the GKP solution the volume direction is a flat direction. The non-vanishing scalar potential for the anti-D3-brane in \eqref{eq:scalarpot}, that is proportional to $1/(vol_6)^\frac43$, would then lead to a runaway for the \K modulus. In order to avoid this, the KKLT scenario~\cite{Kachru:2003aw} includes a non-perturbative correction that can either arise from Euclidean D3-branes or from a gaugino condensate on a stack of D7-branes. The effect of this non-perturbative contribution on the background as well as the anti-D3-brane uplift has recently received considerable attention~\cite{Moritz:2017xto, Moritz:2018sui, Kallosh:2018wme, Moritz:2018ani, Kallosh:2018psh, Gautason:2018gln, Hamada:2018qef, Kallosh:2019axr, Kallosh:2019oxv, Hamada:2019ack, Carta:2019rhx, Gautason:2019jwq}. At the heart of this discussion is the question of whether the gaugino condensation on a stack of D7-branes can be described in ten dimensions and, if that is the case, what the detailed backreaction of the gaugino condensate on the anti-D3-brane is.
When the gaugino condensation or Euclidean D3-brane are taken into account, the KKLT background will have an extra term in the superpotential of the form 
\be
\label{Wnp}
W_{np} = A e^{\rmi aT}\,,
\ee
where $A$ is a function that generically depends on the anti-D3-brane fields. Since the gaugino condensation or Euclidean D3-brane arise from the Calabi-Yau bulk region, while we are studying an anti-D3-brane sitting at the bottom of a highly warped throat, these additional terms are expected to be highly suppressed compared to the tree-level potential in equation~\eqref{eq:scalarpot}. After a lively debate in the recent literature~\cite{Moritz:2017xto, Moritz:2018sui, Kallosh:2018wme, Moritz:2018ani, Kallosh:2018psh, Gautason:2018gln, Hamada:2018qef, Kallosh:2019axr, Kallosh:2019oxv, Hamada:2019ack, Carta:2019rhx, Gautason:2019jwq}, there seems to be some consensus that this is indeed the case. For this reason we will neglect these corrections here, which is in some sense generic.\footnote{For a supersymmetric D3-brane the tree-level scalar potential vanishes and these terms provide a very interesting, small potential that was first studied in~\cite{Baumann:2007ah}.} However, many throats, like for example the Klebanov-Strassler geometry~\cite{Klebanov:2000hb}, have isometries so that the scalar potential in equation \eqref{eq:scalarpot} can have flat directions that would be lifted by higher order corrections. Such higher order corrections could arise in particular setups from the superpotential in equation \eqref{Wnp}. Such light moduli arising from the anti-D3-brane were first studied in \cite{Aharony:2005ez} and it would be interesting to study this in more detail for concrete setups using for example the tools developed in \cite{Gandhi:2011id}.

\subsection{The fermionic action}
The fermionic part of the action plays a crucial role in understanding the low energy effective description of the anti-D3-brane in the GKP or KKLT background. The reason for this is that the anti-D3-brane breaks supersymmetry spontaneously and one (combination) of the world volume fermions has to be the Goldstino. As we will explain in the next section, this Goldstino can be described in terms of a nilpotent chiral multiplet that couples to the standard four-dimensional $\N=1$ supergravity theory one obtains from the closed string sector.

How the anti-D3-brane provides the Goldstino is not straightforward and requires some explanation. We will therefore comment on this before actually presenting the action.
The fermionic action for a Dp-brane in a flux background is only known to quadratic order in the fermions~\cite{Grana:2002tu, Grana:2003ek, Marolf:2003ye, Tripathy:2005hv, Martucci:2005rb, Bergshoeff:2013pia} and the anti-D3-brane has been studied in this context in~\cite{McGuirk:2012sb, Kallosh:2014wsa, Bergshoeff:2015jxa, GarciadelMoral:2017vnz}. In particular, the four world volume fermions on the anti-D3-brane can be divided into $\lambda$, which is a singlet under the SU$(3)$ holonomy group of the internal manifold, and $\chi^i$, with $i=1,2,3,$ which transform as a triplet. The masses and some of the couplings of these fermions are controlled by the imaginary self dual (ISD) part
\be
G_3^{\rm ISD} = \frac12 (G_3 - \rmi *_6 G_3)
\ee
of the background flux $G_3=F_3-\rmi e^{-\phi}H_3$~\cite{McGuirk:2012sb, Bergshoeff:2015jxa}, where in our conventions $F_3 = dC_2 - C_0 H_3$. In particular, the mass of the singlet arises from flux of (0,3) type, the interaction between $\lambda$ and the $\chi^i$ is proportional to non-primitive $(1,2)$ flux and the masses of the $\chi^i$ are determined by primitive (2,1) flux. In a supersymmetric GKP background, $\lambda$ is the Goldstino and correspondingly it does not mix with the $\chi^i$ and it has no mass term. This is consistent with the fact that the anti-D3-brane is the only source of supersymmetry breaking in this context~\cite{Bergshoeff:2015jxa}.

In  a non-supersymmetric GKP background we have a non-vanishing Gukov--Vafa--Witten superpotential~\cite{Gukov:1999ya}
\be
\label{WGVW}
W_{GVW} =\int G_3 \w  \Omega \neq 0\,,
\ee
where $\Omega$ is the (3,0) form of the Calabi-Yau, and therefore we get a non-vanishing F-term for the K\"ahler modulus $T$
\be
D_T W_{GVW} = K_T W_{GVW} \neq 0\,.
\ee
As a consequence, in a non-supersymmetric GKP background with~\eqref{WGVW}, the $G_3$ flux must contain a non-vanishing (0,3) piece and the background itself breaks supersymmetry spontaneously. In such a situation, the Goldstino is a closed string fermion. If we add an anti-D3-brane to this background, then the singlet $\lambda$ will also have a mass and cannot (and does not have to) be the Goldstino. Indeed, since both the anti-D3-brane as well as the background (0,3) $G_3$ flux break both supersymmetry spontaneously, the Goldstino is expected to be a linear combination of $\lambda$ and of the closed string fermion which was the Goldstino before the addition of the anti-D3-brane.

The actual KKLT background of interest has, besides a non-vanishing $W_{GVW}$, a non-perturbative superpotential term given in \eqref{Wnp}. In this case one can find a supersymmetric solution
\be
D_T (W_{GVW} + W_{np})= 0\,, 
\ee
which gives $\partial_T W_{np} = - K_T (W_{GVW}+ W_{np})$. This solution is a supersymmetric AdS vacuum that is uplifted by the anti-D3-brane to the KKLT dS vacuum. At this point, however, one might have noticed the following issue: the anti-D3-brane is actually the sole source of supersymmetry breaking and therefore it needs to provide the massless Goldstino. On the other hand, the background has (2,1) as well as (0,3) $G_3$ flux, which seem to give a mass to $\chi^i$ and to $\lambda$ as well~\cite{Bergshoeff:2015jxa}. In other words, all fermions appear to be massive and it is not clear whether a massless Goldstino is present. The resolution of this apparent puzzle is due to the behavior of the (0,3) $G_3$ flux in the presence of a gaugino condensate on a stack of D7-branes. In particular, it was shown in~\cite{Baumann:2010sx, Dymarsky:2010mf} that the (0,3) $G_3$ flux localizes on top of the D7-branes that are located in the bulk of the warped Calabi-Yau manifold. Therefore, the pull-back of this (0,3) $G_3$ flux onto the anti-D3-brane world volume vanishes, since the anti-D3-branes sits at the bottom of a warped throat. Thus, $\lambda$ does not get a mass and is the Goldstino provided by the anti-D3-brane, which is the sole source of supersymmetry breaking.\footnote{This is only true in the strict probe limit. Once the anti-D3-brane backreacts onto the geometry via the uplift term $V_{\rm up }\propto 1/(-\rmi (T-\bar{T}))^2$, the $T$ modulus shifts away from the supersymmetric minimum. As a consequence, $D_T W$ will not be zero anymore and the Goldstino will be a mixture of $\lambda$ and the fermionic partner of the $T$ modulus.}

Having clarified how the anti-D3-brane action in the KKLT background provides the massless Goldstino plus three more massive fermions, we now work out the couplings of these fermions to the closed string moduli $\tau$, $T$ and $U^A$.
The fermionic anti-D3-brane action was studied for example in~\cite{Grana:2002tu,McGuirk:2012sb,Bergshoeff:2015jxa,GarciadelMoral:2017vnz}. The part of the action that is quadratic in worldsheet fermions is given in Einstein frame by~\cite{Martucci:2005rb, Bergshoeff:2005yp, Bergshoeff:2015jxa}\footnote{This can be obtained from the string frame result in~\cite{Bergshoeff:2005yp, Bergshoeff:2015jxa} with: $g^{\rm S}_{\mu\nu}= e^{\frac{\phi}{2} + 2\A - 4 u} g_{\mu\nu}$ and $\theta^{\rm S} = e^{\frac{\phi}{8}-u+\frac{\A}{2}}\theta$. We follow the conventions of~\cite{Bergshoeff:2015jxa}, which uses the action given in equations (4.4) to (4.6) of~\cite{Bergshoeff:2005yp}. We re-derived the coupling to $F_1$ starting from equations (A.5) and (A.7) in~\cite{Bergshoeff:2005yp} and found the opposite sign for the corresponding term. Our expression agrees with the generic four dimensional supergravity action as we check below.}
\be
\begin{aligned}
\label{Sfer10d}
S^{\Db}_{fer} = 2\int &d^4x \sqrt{-g_4}\, \bigg[ e^{4\mathcal{A}-8 u}\bar\theta\Gamma^\mu \left(\nabla_\mu
-\frac{1}{4}e^\phi F_\mu \tilde \Gamma_{0123}\right)\theta\cr
&+\frac{1}{8\cdot 4!} \bar \theta\left(e^{\frac{16\A}{3}-\frac{32u}{3}}\Gamma^{\mu mnpq}\, F_{\mu mnpq}-2e^{\frac{8\A}{3}-\frac{16u}{3}} \Gamma^{\mu\nu\rho mn} \,F_{\mu\nu\rho mn}\right)\tilde \Gamma_{0123}\theta\cr
&-\frac{\rmi}{24}e^{6\mathcal{A}-12u + \frac{\phi}{2}}\,(G_{mnp}^{\rm ISD}-\bar G_{mnp}^{\rm ISD})\,\bar\theta\Gamma^{mnp}\theta\bigg], 
\end{aligned}
\ee
where $\theta$ is a 16-component Majorana-Weyl spinor of type IIB theory, the indices $m,n,\ldots =4,5,\dots,9$ are internal, $\Gamma_{\ldots}$ has curved but unwarped indices and $\tilde \Gamma_{\ldots}$ has flat indices. Following~\cite{Grana:2002tu,Bergshoeff:2015jxa}, we can decompose the spinor $\theta$ into four-dimensional Weyl spinors. In the notation of \cite{freedman2012supergravity}, we have an SU$(3)$ singlet  $P_L\lambda$ and a triplet $P_L \chi^i$. The reduction of the last line and the kinetic term of \eqref{Sfer10d} was performed in detail in \cite{Bergshoeff:2015jxa}. If the brane action is evaluated on a fixed background the remaining terms vanish. On the contrary, they should be taken into account when the background fields are dynamical, as it is the case in our setup. To the best of our knowledge, these terms are worked out here for the first time.

We start by considering the contribution arising from the spin connection. In particular, we have
\be
\bar\theta \Gamma^\mu \nabla_\mu \theta = \bar\theta \Gamma^\mu \lp \partial_\mu +\frac14 \omega_\mu^{\tilde a \tilde b} \tilde\Gamma_{\tilde a\tilde b}+\frac14 \omega_\mu^{i\ib} \tilde\Gamma_{i\ib} \rp\theta\,,
\ee
where the flat indices take values $\tilde a,\tilde b=0,1,2,3$ and $i,\ib=1,2,3$. Since the spin connection with mixed indices vanishes, $\omega_\mu^{\tilde a i}=\omega_\mu^{\tilde a \ib}=0$, we have already omitted the corresponding terms. The first two terms in the equation above construct the 4d covariant derivative, while the last term gives rise to a coupling between the fermions and the complex structure moduli, which can be calculated as follows. In our setup the metric is block diagonal with a 4d and 6d part and therefore the vielbein is likewise block diagonal. The six dimensional internal vielbein part satisfies then $e^a_i g_{a\bar b}e^{\bar b}_\ib  = \delta_{i\ib}$, so that $e^a_i$ is a function of the \K modulus $T$ and of the complex structure moduli $U^A$. Since the holomorphic (3,0)-form $\Omega_{abc} =\epsilon_{ijk}e^i_a e^j_b e^k_c$ only depends on the $U^A$ and not on the $\bar U^{A}$, we can conclude that $e^i_a$ and its inverse $e^a_i$ depend only on $T$, $\bar T$ and the $U^A$ but not on $\bar U^A$. Namely, we have
\be
\partial_\mu e^a_i = \lp \partial_T e^a_i \rp \partial_\mu T + (\partial_{\bar T} e^a_i) \partial_\mu \bar T  + \lp \partial_{U^A} e^a_i\rp \partial_\mu U^A\,.
\ee
Having a single volume modulus, the metric and the vielbein have a simple overall volume dependence, i.e. they depend to leading order in large volume on $(T-\bar T)$ to some power, so that $\partial_T e^a_i =- \partial_{\bar T} e^a_i \propto e^a_i$. This means that the relevant spin connection reduces to
\be\label{eq:spinconnection}
\omega_\mu^{i\ib} =   e^{\bar a i}\partial_\mu e_{\bar a}^\ib  - e^{a \ib} \partial_\mu e_a^i= e^{\bar a i}(\partial_{\bar U^A} e_{\bar a}^\ib) \partial_\mu \bar U^A-e^{a \ib} (\partial_{U^A} e_a^i) \partial_\mu U^A \,.
\ee
We will now use the fact that we have only a single \K modulus and correspondingly only a single (1,1)-form to simplify the above expression substantially. In particular, from the spin connection we can define a 2-form that has to be proportional to the \K form $J$, or in flat indices to $\delta_{i\ib}$, if it is in cohomology \footnote{It is not clear to us that this 2-form indeed has to be in cohomology for generic CY$_3$ manifolds. For toroidal orbifold examples this is indeed the case, but we lack a generic argument. The following is therefore not a strict mathematical proof. We thank Harald Skarke for discussing this point.}
\be
\omega_{\mu\, i\ib}\, e^i_a e^{\ib}_{\bar b} \propto \rmi J_{a\bar b} = \delta_{i\ib} e^i_a e^{\ib}_{\bar b}\,.
\ee
We can then write 
\be
\omega_\mu^{i\ib} \tilde\Gamma_{i\ib} = \frac13 \omega_\mu^{i\ib} \delta_{i\ib} \delta^{j\jb} \tilde\Gamma_{j\jb}
\ee
and the first new fermionic contribution to the 4d anti-D3-brane action is
\ba
\bar\theta \Gamma^\mu \omega_\mu^{i\ib} \tilde\Gamma_{i\ib} \theta &=& \frac{1}{3} \bar\theta \Gamma^\mu  \omega_\mu^{i\ib} \delta_{i\ib} \delta^{j\jb} \tilde\Gamma_{j\jb} \theta = \frac{1}{3}\omega_\mu^{k\bar k} \delta_{k\bar k}  \lp 3\bar\lambda P_R\gamma^\mu \lambda - \delta_{i\bar \jmath}\bar\chi^{\bar \jmath} P_R \gamma^\mu \chi^i \rp\\
&=&  \frac{1}{3} \delta_{i\ib} \lp  e^{\bar a i}(\partial_{\bar U^A} e_{\bar a}^\ib) \partial_\mu \bar U^A - e^{a \ib} (\partial_{U^A} e_a^i) \partial_\mu U^A\rp \lp 3\bar\lambda P_R\gamma^\mu \lambda - \delta_{j\bar \jmath}\bar\chi^{\bar \jmath} P_R \gamma^\mu \chi^j \rp\nn\,.
\ea

The second new term involves a coupling to the derivative $F_\mu = \partial_\mu C_0= \partial_\mu {\rm Re} \tau$. Its calculation is simpler with respect to the previous case and it gives directly
\be
e^{\phi}F_\mu \bar\theta \Gamma^\mu\tilde\Gamma_{0123} \theta =\frac{\partial_\mu {\rm Re}(\tau)}{{\rm Im}(\tau)} \bar\theta \Gamma^\mu\tilde\Gamma_{0123} \theta = -\rmi \frac{\partial_\mu {\rm Re}(\tau)}{{\rm Im}(\tau)}  \lp \bar\lambda P_R\gamma^\mu \lambda + \delta_{i\bar \jmath}\bar\chi^{\bar \jmath} P_R \gamma^\mu \chi^i \rp\,.
\ee

The last contribution in \eqref{Sfer10d} that we have to calculate is a derivative coupling to the $C_4$ axion. To this purpose, we recall that for a Calabi-Yau manifold with a single \K modulus $T$, there is only a single $(2,2)$-form that we denote by $Y_{2,2}$ and that is normalized such that it integrates to one on the single 4-cylcle $\Sigma_4$. In particular, the \K modulus is constructed out of the 4-forms $C_4$ and $J\wedge J$ and it has to be holomorphic, since it is described by a chiral multiplet in four dimensions. We can therefore decompose it on the basis given by $Y_{2,2}$ and find
\be
T \equiv \int_{\Sigma_4} \lp C_4 -\frac{\rmi}{2} J\wedge J \rp =  \int_{\Sigma_4} c_4(x^\mu) Y_{2,2} + \rmi {\rm Im}(T)  \int_{\Sigma_4} Y_{2,2}\,.
\ee
From the above we can identify directly $C_4 \supset c_4(x^\mu) Y_{2,2}= -c_4(x^\mu) \frac{ J\w J}{2 {\rm Im}(T)}$. Using that $\rmi e^u_i e^{\bar u}_\ib J_{u\bar u} = \delta_{i\ib}$, where $u,\bar u$ are curved and warped indices, while recalling that the matrix $\Gamma^{\mu npqr}$ appearing in \eqref{Sfer10d} has real curved but unwarped indices, we can calculate finally the desired term
\ba
\frac{1}{4!}e^{\frac{4\A}{3}-\frac{8u}{3}} \bar \theta\Gamma^{\mu npqr} \tilde\Gamma_{0123}\, F_{\mu npqr}\theta &=& \frac{\partial_\mu {\rm Re}(T)}{2{\rm Im}(T)} \bar \theta \delta_{i\ib} \delta_{j\jb} \Gamma^{\mu} \tilde\Gamma^{i \ib j \jb} \tilde\Gamma_{0123}\theta \cr
&=&-\rmi \frac{\partial_\mu {\rm Re}(T)}{{\rm Im}(T)}\lp 3\bar\lambda P_R\gamma^\mu \lambda - \delta_{i\bar \jmath}\bar\chi^{\bar \jmath} P_R \gamma^\mu \chi^i \rp\,.
\ea
Actually, this is only one of the two contributions containing $C_4$ and appearing in \eqref{Sfer10d}. However, due to the self-duality of $dC_4$ in ten dimensions, one finds that the four-dimensional 2-form $C_4 \supset c_{4,\mu\nu}^{(2)} dx^\mu \w dx^\nu\w Y_{1,1}$ is dual in four dimensions to $c_4(x^\mu)$, so that the two corresponding terms in the 3-brane action are equal, namely
\be
F_{\mu npqr} \Gamma^{\mu npqr} = -2 e^{-\frac{8\A}{3}+\frac{16u}{3}} F_{\mu\nu\rho mn} \Gamma^{\mu\nu\rho mn}\tilde\Gamma_{*}\,,
\ee
and therefore the two terms in \eqref{Sfer10d} add up.

Up to total derivatives and written in terms of four-dimensional spinors the action \eqref{Sfer10d} becomes then
\begin{align}
S^{\Db}_{fer}  &=2 \int d^4 x \sqrt{-g_4}e^{4\mathcal{A}-8 u}\bigg[\bar\lambda P_R\gamma^\mu \nabla_\mu \lambda + \delta_{i\bar \jmath}\bar\chi^{\bar \jmath} P_R \gamma^\mu \nabla_\mu \chi^i \nonumber\\
&\qquad\qquad\qquad\qquad\qquad+\frac{\rmi}{4 {\rm Im}(\tau)} \partial_\mu {\rm Re}(\tau)\lp \bar\lambda P_R\gamma^\mu \lambda + \delta_{i\bar \jmath}\bar\chi^{\bar \jmath} P_R \gamma^\mu \chi^i \rp\nn\\
&\qquad\qquad\qquad\qquad\qquad-\frac{\rmi}{4 {\rm Im}(T)} \partial_\mu {\rm Re}(T)\lp 3\bar\lambda P_R\gamma^\mu \lambda - \delta_{i\bar \jmath}\bar\chi^{\bar \jmath} P_R \gamma^\mu \chi^i \rp\\
&\qquad\qquad\qquad\qquad\qquad+\frac{1}{12} \omega_\mu^{k\bar k} \delta_{k\bar k}\lp 3\bar\lambda P_R\gamma^\mu \lambda - \delta_{i\bar \jmath}\bar\chi^{\bar \jmath} P_R \gamma^\mu \chi^i \rp\nn\\
&\qquad\qquad\qquad\qquad\qquad+\left(\frac12 m \bar\lambda P_L\lambda +m_i\bar\lambda P_L\chi^i + \frac12 m_{ij} \bar\chi^i P_L \chi^j+c.c.\right)\bigg]\nonumber\,.
\end{align}
The masses depend on the $G^{\rm ISD}_3$ flux as
\begin{align}
m &= \frac{\sqrt 2}{12}\rmi e^{2\A-4u+\frac{\phi}{2}}\bar \Omega^{abc}\bar G_{abc}^{\rm ISD}\,,\\
m_i &= -\frac{\sqrt 2}{4}e^{2\A-4u+\frac{\phi}{2}}e^a_i \bar G_{ab\bar c}^{\rm ISD}J^{b\bar c}\,,\\
\label{mij}
m_{ij} &= \frac{\sqrt 2}{8} \rmi e^{2\A-4u+\frac{\phi}{2}}(e^c_i e^d_j+e^c_j e^d_i)\Omega_{abc} g^{a\bar a}g^{b\bar b}\bar G^{\rm ISD}_{d\bar a\bar b}\,.
\end{align}
As we discussed at the beginning of this subsection, in a GKP background or in the KKLT setup with gaugino condensation on a stack of D7-branes in the bulk that is away from the anti-D3-brane, the pullback of the (0,3) part of the $G_3$ flux onto the anti-D3-brane world volume vanishes. In addition, in the same background the (2,1) part of $G_3^{\rm ISD}$ is primitive, namely $G_{ab\bar c}^{\rm ISD}J^{b\bar c}=0$, where $J_{a\bar b}$ is the \K form on the Calabi-Yau. As a consequence of these two facts we have that 
\be
m = 0 =m_i
\ee
and then the singlet fermion $\lambda$ remains massless, while the $\chi^i$ generically have non-vanishing mass terms. $\lambda$ will therefore be the Goldstino associated to the broken supersymmetry. It arises from the brane, as is expected for an anti-D3-brane added to a supersymmetric background.

The complete anti-D3-brane action in the KKLT background is finally
\be
\begin{aligned}
\label{SDbcomplete}
S^{\Db} &= S^{\Db}_{bos} + S^{\Db}_{fer}\\
&=- \int d^4x \sqrt{-g_4}\bigg(2  e^{4 \A-8 u}+\frac12 e^{\frac43 \A-\frac83 u}g_{a\bar b}\partial_\mu H^a \partial^\mu \bar H^{\bar b} \\
&\quad\qquad\qquad\qquad\quad+ \frac{{\rm Im}(\tau)}{4}F_{\mu\nu}F^{\mu\nu} - \frac{{\rm Re}(\tau)}{8}\frac{\epsilon^{\mu\nu\rho\sigma}}{\sqrt{-g_4}}F_{\mu\nu}F_{\rho\sigma}\bigg)\\
&+2 \int d^4x\sqrt{-g_4}e^{4\mathcal{A}-8 u}\bigg[\bar\lambda P_R\gamma^\mu \nabla_\mu \lambda + \delta_{i\bar \jmath}\bar\chi^{\bar \jmath} P_R \gamma^\mu \nabla_\mu \chi^i  \\
&\qquad\qquad\qquad\qquad\qquad+\frac{\rmi}{4 {\rm Im}(\tau)} \partial_\mu {\rm Re}(\tau)\lp \bar\lambda P_R\gamma^\mu \lambda + \delta_{i\bar \jmath}\bar\chi^{\bar \jmath} P_R \gamma^\mu \chi^i \rp\\
&\qquad\qquad\qquad\qquad\qquad-\frac{\rmi}{4 {\rm Im}(T)} \partial_\mu {\rm Re}(T)\lp 3\bar\lambda P_R\gamma^\mu \lambda - \delta_{i\bar \jmath}\bar\chi^{\bar \jmath} P_R \gamma^\mu \chi^i \rp\\
&\qquad\qquad\qquad\qquad\qquad+\frac{1}{12} \omega_\mu^{k\bar k} \delta_{k\bar k}\lp 3\bar\lambda P_R\gamma^\mu \lambda - \delta_{i\bar \jmath}\bar\chi^{\bar \jmath} P_R \gamma^\mu \chi^i \rp\\
&\qquad\qquad\qquad\qquad\qquad+\frac12 m_{ij} \bar\chi^i P_L \chi^j+\frac12 \overline m_{\bar \imath \bar \jmath} \bar\chi^{\bar \imath} P_R \chi^{\bar\jmath}\bigg]\,.
\end{aligned}
\ee

\subsection{The supersymmetric D3-brane action}\label{sec:D3K}
The above action is the leading order component action for an anti-D3-brane in the KKLT or GKP background and we will show how to rewrite it in terms of $\N=1$ supergravity in section \ref{sec:sugraaction}. However, before doing so, it is instructive to perform a simple check on our result and compare it to the known couplings in the D3-brane action. We devote therefore the present subsection to this purpose.

Recall that the D3-brane differs from the anti-D3-brane action by a sign flip of the RR-fields. In the supersymmetric D3-brane case one also has to take into account that the first `uplift term' in the bosonic action and the fermionic mass terms vanish. What remains is then a standard $\N=1$ supergravity action for a single vector multiplet, containing $\lambda$ and $A_\mu$, and three chiral multiplets, containing $H^a$ and $\chi^a \equiv e^a_i \chi^i$. Explicitly, it is given by

\begin{align}
\label{SDcomplete}
S^{D3} &= S^{D3}_{bos} + S^{D3}_{fer}\nn\\
&=- \int d^4x \sqrt{-g_4}\bigg(\frac12 e^{\frac43 \A-\frac83 u}g_{a\bar b}\partial_\mu H^a \partial^\mu \bar H^{\bar b} \nn\\
&\quad\qquad\qquad\qquad\quad+ \frac{{\rm Im}(\tau)}{4}F_{\mu\nu}F^{\mu\nu} + \frac{{\rm Re}(\tau)}{8}\frac{\epsilon^{\mu\nu\rho\sigma}}{\sqrt{-g_4}}F_{\mu\nu}F_{\rho\sigma}\bigg)\nn\\
&+2 \int d^4x\sqrt{-g_4}e^{4\mathcal{A}-8 u}\bigg[\bar\lambda P_R\gamma^\mu \nabla_\mu \lambda + \delta_{i\bar \jmath}\bar\chi^{\bar \jmath} P_R \gamma^\mu \nabla_\mu \chi^i  \\
&\qquad\qquad\qquad\qquad\qquad-\frac{\rmi}{4 {\rm Im}(\tau)} \partial_\mu {\rm Re}(\tau)\lp \bar\lambda P_R\gamma^\mu \lambda + \delta_{i\bar \jmath}\bar\chi^{\bar \jmath} P_R \gamma^\mu \chi^i \rp\nn\\
&\qquad\qquad\qquad\qquad\qquad+\frac{\rmi}{4 {\rm Im}(T)} \partial_\mu {\rm Re}(T)\lp 3\bar\lambda P_R\gamma^\mu \lambda - \delta_{i\bar \jmath}\bar\chi^{\bar \jmath} P_R \gamma^\mu \chi^i \rp\nn\\
&\qquad\qquad\qquad\qquad\qquad+\frac{1}{12} \omega_\mu^{k\bar k} \delta_{k\bar k}\lp 3\bar\lambda P_R\gamma^\mu \lambda - \delta_{i\bar \jmath}\bar\chi^{\bar \jmath} P_R \gamma^\mu \chi^i \rp\bigg]\nn\,.
\end{align}

It is interesting to study the derivative couplings involving (derivatives of) the closed string axions and the open string fermions on the supersymmetric D3-brane, since this will provide useful information about the form of the \K potential even for the anti-D3-brane. Let us begin with the coupling to $\partial_\mu {\rm \tau}$. The fermions $\chi^i$ come in chiral multiplets and carry no U$(1)$ charge, but via their \K covariant derivative they couple to all scalars. In particular, they couple to $\tau$ via terms (see for example chapter 18 in \cite{freedman2012supergravity})
\ba
\L^{\rm SUGRA} &\supset&  -\delta_{i\jb} \bar\chi^\jb P_R \gamma^\mu \lp \partial_\mu -\frac14\ls\partial_\mu \tau \partial_\tau K - \partial_\mu \bar \tau \partial_{\bar \tau} K\rs\rp \chi^i \cr 
&&-\frac12\delta_{i\jb} \bar\chi^\jb P_R \gamma^\mu \Gamma^i_{k \tau} \partial_\mu \tau \chi^k -\frac12\delta_{i\jb} \bar\chi^i P_L \gamma^\mu \Gamma^\jb_{\bar k \bar\tau} \partial_\mu \bar\tau \chi^{\bar k} \,.\ea
We find that the \K potential $K^{(\tau)}=-\log\ls-\rmi (\tau-\bar{\tau})\rs$ leads precisely to the coupling involving $\partial_\mu {\rm Re}(\tau)/{\rm Im}(\tau)$ that is given in \eqref{SDcomplete}, via the two terms in square brackets above. In particular, the prefactor of this term relative to the kinetic term also agrees with the generic supergravity result. This means that the mixed Christoffel symbols $\Gamma^i_{j \tau}$ have to vanish, which is indeed the case for appropriately chosen \K potentials, like the one given in equation \eqref{eq:Kpot} that couples the scalars in the chiral multiplets only to $T$ and not to $\tau$.

The couplings of $\tau$ and $\lambda$ can be likewise read off from the general supergravity action. Recall that the gauge kinetic function for a D3-brane is $f(\tau) =-\rmi \tau$. The kinetic term for the gaugino is normalized so that its prefactor is Re$(f)={\rm Im}(\tau) = e^{-\phi}$, so we have to rescale $\lambda = e^{-\frac\phi2} \lambda'$, but this does not lead to new derivative terms since $\bar\lambda \gamma^\mu\lambda=0$ for Majorana spinors in four dimensions. The standard supergravity action for $\lambda'$ contains then terms of the form
\be
\L^{\rm SUGRA} \supset-\frac12  {\rm Re}(f) \bar \lambda' \gamma^\mu\lp \partial_\mu + \frac14 \lp\partial_\mu \tau \partial_\tau K - \partial_\mu \bar \tau \partial_{\bar \tau} K\rp \gamma_* \rp \lambda' + \frac{\rmi}{4} \partial_\mu {\rm Im}(f) \bar \lambda' \gamma_* \gamma^\mu \lambda'\,.
\ee
These combine correctly to give the coupling of $\lambda$ to $\partial_\mu {\rm Re}(\tau)/{\rm Im}(\tau)$ that is given in \eqref{SDcomplete}.

We now proceed and look at the coupling to $\partial_\mu {\rm Re}(T)$ by performing a similar analysis. For the $\chi^i$ that are in chiral multiplets, these couplings arise again from the standard supergravity terms
\ba
\L^{\rm SUGRA} &\supset&  -\delta_{i\jb} \bar\chi^\jb P_R \gamma^\mu \lp \partial_\mu -\frac14\ls\partial_\mu T \partial_T K - \partial_\mu \bar T \partial_{\bar T} K\rs\rp \chi^i \cr 
&&- \frac12\delta_{i\jb} \bar\chi^\jb P_R \gamma^\mu \Gamma^i_{k T} \partial_\mu T \chi^k -\frac12\delta_{i\jb} \bar\chi^i P_L \gamma^\mu \Gamma^\jb_{\bar k \bar T} \partial_\mu \bar T \chi^{\bar k} \,.
\ea
This time, however, the Christoffel symbols are not vanishing. Indeed, we find for the \K potential in equation \eqref{eq:Kpot} that $\Gamma^i_{j T} \approx \frac{\rmi}{2{\rm Im}(T)}$ in the large volume limit, where we neglect terms involving $k(H,\bar H)$ compared to Im$(T)$. These Christoffel symbols combine then with the terms involving partial derivatives of the \K potential $K$ to give again the terms in the component action \eqref{SDcomplete} with the correct coefficient.

Since for the gauginos $\lambda$ there is no contribution involving the Christoffel symbols, the standard supergravity action has terms, written using $\lambda'=e^{\frac{\phi}{2}} \lambda$, that are simply
\be
\L^{\rm SUGRA} \supset-\frac12  {\rm Re}(f) \bar \lambda' \gamma^\mu\lp \partial_\mu + \frac14 \lp\partial_\mu T \partial_T K - \partial_\mu \bar T \partial_{\bar T} K\rp \gamma_* \rp \lambda' \,.
\ee
They match again with the component action \eqref{SDcomplete} and the very absence of the Christoffel terms for $\lambda$ compared to $\chi^i$ explains the different prefactor for the corresponding terms.

Last and most interestingly, we look at the terms involving the complex structure moduli $U^A$. The two terms in the spin connection in equation \eqref{eq:spinconnection} are independent, since one is proportional to $\partial_\mu U^A$ and the other to $\partial_\mu \bar U^A$. Following a reasoning similar to that of the previous subsection, this means these terms have to be both proportional to $\delta_{i\ib}$. Therefore, using that $\partial_\mu (e_u^i e^{u\ib})=0$, we can rewrite
\be
\partial_{U^A} e_u^i =-e^i_v(\partial_{U^A} e^{v\ib}) e_{u\ib} = -\frac13 e^j_v(\partial_{U^A} e^{v\jb}) \delta_{j\jb} \delta^{i\ib} e_{u\ib} = -\frac13 e^j_v(\partial_{U^A} e^{v\jb}) \delta_{j\jb} \, e^i_{u}\,,
\ee
where $u,v,\dots=1,2,3$ are curved warped indices.
This expression can be rephrased in terms of the holomorphic (3,0)-form $\Omega$ on the Calabi-Yau manifold by using that
\ba\label{eq:Uderivative}
\partial_{U^A} \Omega &=&\frac{1}{3!\cdot 3!} \partial_{U^A} ( e^i_u e^j_v e^k_w \epsilon_{ijk} dz^u\w dz^v \w dz^w) \cr
&=& \frac{3}{3!\cdot 3!} (\partial_{U^A} e^i_u) e^j_v e^k_w \epsilon_{ijk} dz^u\w dz^v \w dz^w\cr
&=& -\frac{1}{3!\cdot 3!} e^l_t(\partial_{U^A} e^{t\jb}) \delta_{l\jb} \, e^i_{u} e^j_v e^k_w \epsilon_{ijk} dz^u\w dz^v \w dz^w\cr
&=& - e^j_v(\partial_{U^A} e^{v\jb}) \delta_{j\jb} \, \Omega\,.
\ea
On the other hand, we can expand $\Omega$ on a cohomology basis, namely $\Omega=Z^K \alpha_K - F_K \beta^K$, where $Z_K$ and $F^K$ are functions of the $U^A$, while the $\alpha_K$ and $\beta^K$ are a basis for the 3-forms with the only non-vanishing integrals $\int \alpha_K \w \beta^L = \delta_K^L$. In particular, it follows from equation \eqref{eq:Uderivative} that
\be
\partial_{U^A} Z^K = - e^j_v(\partial_{U^A} e^{v\jb}) \delta_{j\jb} Z^K\,, \qquad \partial_{U^A} F_K = - e^j_v(\partial_{U^A} e^{v\jb}) \delta_{j\jb} F_K\,.
\ee
We introduce now the \K potential for the complex structure moduli
\be
K^{(U)} = - \log \ls -\rmi \int \Omega \w \bar \Omega \rs = - \log \ls \rmi \lp Z^K\bar F_K-\bar Z^K F_K \rp\rs\,,
\ee
and we can use this to find eventually
\ba
\omega_\mu^{i\ib} \delta_{i\ib} &=& \delta_{i\ib}\lp  e^{\bar a i}(\partial_{\bar U^A} e_{\bar a}^\ib) \partial_\mu \bar U^A -e^{a \ib} (\partial_{U^A} e_a^i) \partial_\mu U^A\rp\cr
 &=& \partial_{U^A} K^{(U)} \partial_\mu U^A-\partial_{\bar U^A} K^{(U)} \partial_\mu \bar U^A \,.
\ea
With this result at our disposal, we can again match the component action for the D3-brane with the standard supergravity expression. It is essential to notice that in the component action \eqref{SDcomplete} the couplings of $\lambda$ and $\chi^i$ to $\partial_\mu {\rm Re}(T)$ and to $\partial_\mu {\rm Re}(U^A)$ (via the spin connection term), have the same numerical coefficient.
As a consequence, the complex structure sector and the \K sector have to couple to the chiral multiplets on the D3-brane in the same way. This observation leads us to propose the \K potential 
\ba\label{eq:D3Kahler}
K &=& -\log\ls-\rmi (\tau-\bar{\tau})\rs -3\log\ls-\rmi (T-\bar{T})\lp-\rmi \int \Omega \w \bar \Omega \rp^{\frac13} + k(H,\bar H)\rs\cr
&=& -\log\ls-\rmi (\tau-\bar{\tau})\rs -\log\ls-\rmi \int \Omega \w \bar \Omega\rs \cr
&&\qquad\qquad\qquad\qquad-3\log\ls-\rmi (T-\bar{T}) + \frac{k(H,\bar H)}{\lp-\rmi \int \Omega \w \bar \Omega \rp^{\frac13}}\rs\,,
\ea
which produces indeed the couplings to $\partial_\mu {\rm Im}(\tau)$ and $\partial_\mu {\rm Im}(T)$ as discussed above. In addition, it gives the correct coupling to $\chi^i$ via the standard supergravity terms
\ba
\L^{\rm SUGRA} &\supset&  -\delta_{i\jb} \bar\chi^\jb P_R \gamma^\mu \lp \partial_\mu -\frac14\ls\partial_\mu U^A \partial_{U^A} K - \partial_\mu \bar U^A \partial_{\bar U^A} K\rs\rp \chi^i \cr 
&&- \frac12\delta_{i\jb} \bar\chi^\jb P_R \gamma^\mu \Gamma^i_{k U^A} \partial_\mu U^A \chi^k -\frac12\delta_{i\jb} \bar\chi^i P_L \gamma^\mu \Gamma^\jb_{\bar k \bar U^A} \partial_\mu \bar U^A \chi^{\bar k} \,,
\ea
if we drop again terms involving $k(H,\bar H)$ and its derivatives, while the couplings to $\lambda'$ can be obtained from
\be
\L^{\rm SUGRA} \supset-\frac12  {\rm Re}(f) \bar \lambda' \gamma^\mu\lp \partial_\mu + \frac14 \lp\partial_\mu U^A \partial_{U^A} K - \partial_\mu \bar U^A \partial_{\bar U^A} K\rp \gamma_* \rp \lambda' \,.
\ee
To conclude, notice that the maybe naively expected \K potential 
\be
K = -\log\ls-\rmi (\tau-\bar{\tau})\rs -\log\ls-\rmi \int \Omega \w \bar \Omega\rs -3\log\ls- \rmi(T-\bar{T})\ + k(H,\bar H)\rs\,,
\ee
\emph{does not} seem to reproduce the component action for the supersymmetric D3-brane. Instead, we need to use the \K potential given above in equation \eqref{eq:D3Kahler}, which couples the world volume scalars $H^a$ to the complex structure moduli $U^A$. Finally, in order to still reproduce the kinetic term for the $H^a$, $k(H,\bar H)$ needs now to be chosen such that $\partial_{H^a}\partial_{\bar{H}^{\bar b}} k(H,\bar H) \approx \frac{1}{6} e^{\frac43 (\A+u)}\lp-\rmi \int \Omega \w \bar \Omega \rp^{\frac13}g_{a\bar b}$.

\bigskip
This concludes our analysis of the (anti-)D3-brane action from the string theory perspective. In section~\ref{sec:sugraaction} we will show how to obtain the anti-D3-brane action in equation \eqref{SDbcomplete} from $\N=1$ supergravity in four dimensions. For this purpose, non-linear (local) supersymmetry will be employed in the language of constrained multiplets. We review the necessary ingredients in the following section.

\section{Constrained multiplets in supergravity}\label{sec:sugracomponent}
In this section we review some ingredients of non-linear realizations and constrained multiplets in supergravity. We focus on a particular set of constraints that we are going to employ in order to describe the anti-D3-brane action~\cite{Vercnocke:2016fbt, Kallosh:2016aep}. It is important to keep in mind, however, that the choice of the required constraints in general is not unique, but different possibilities can occur. See for example \cite{Bandos:2016xyu, GarciadelMoral:2017vnz} for a discussion about this fact. 

We use the conventions of~\cite{freedman2012supergravity}, where 4d fermions are described by four-component Majorana spinors. In the appendix \ref{appA:superspace}, the relevant formulae are given also in the flat superspace language, following the conventions of~\cite{Wess:1992cp}.

\subsection{The Goldstino in a flat background}

When supersymmetry is spontaneously broken a Goldstino is present in the spectrum and transforms non-homogeneously under supersymmetry transformations. A minimal action describing the Goldstino was proposed by Volkov and Akulov~\cite{Volkov:1973ix} and it is of the type
\be 
\label{eq:VAaction}
S_{VA} = - M^4\int E^0 \wedge E^1 \wedge E^2 \wedge E^3, \qquad \text{with} \qquad E^\mu = dx^\mu + \bar{\lambda} \gamma^\mu d\lambda\, ,
\ee
where $\lambda$ is the spin-1/2 Goldstino and $M$ is a parameter of mass dimension one, related to the supersymmetry breaking scale. This action is invariant under the non-linear transformation ($M=1$)
\be 
\delta_{\epsilon} \lambda = \epsilon + \left(\bar{\lambda} \gamma^\mu \epsilon \right) \partial_\mu \lambda\,,
\ee
which closes onto the $\mathcal{N}=1$ supersymmetry algebra.

We now discuss briefly how to reformulate the Volkov--Akulov model in a language in which supersymmetry becomes manifest. When dealing with supersymmetric theories, it is convenient to embed fields into multiplets or superfields. A simple choice consists in identifying the Goldstino with the fermion $P_L\Omega$ of a chiral multiplet\footnote{We denote multiplets on which supersymmetry acts linearly with the same letter as their lowest components.}
\be
X = \left\{X, P_L\Omega,F\right\}.
\ee
This multiplet, however, contains also a scalar $X$ which is not present in the Volkov--Akulov model~\eqref{eq:VAaction}. It is possible to eliminate this scalar in a supersymmetric way by imposing an additional constraint on the multiplet. If we require $X$ to be nilpotent~\cite{Rocek:1978nb, Lindstrom:1979kq, Casalbuoni:1988xh, Komargodski:2009rz}, then the scalar in the lowest component becomes a function of the Goldstino and of the auxiliary field $F$:
\be
\label{X2=0}
X^2 =0 \quad \Leftrightarrow \quad X = \left\{ \frac{\bar \Omega P_L \Omega}{2F}, P_L \Omega, F\right\}.
\ee
An invariant action for this nilpotent chiral multiplet is given by
\begin{equation}
\label{LVAKS}
\begin{aligned}
S&= [X\bar X]_D + M^2[X]_F\\
&= \int d^4 x\left( - \bar \Omega P_L \slashed{\partial} \Omega + \frac{\bar \Omega P_L\Omega}{2F}\Box  \frac{\bar \Omega P_R\Omega}{2\bar F}+ F \bar F + M^2(F+\bar F)\right).
\end{aligned}
\end{equation}
Notice that the equations of motion of the auxiliary fields are modified with respect to the case in which supersymmetry is linearly realized, since the sGoldstino is replaced by a composite expression containing $F$. This is a general feature of models with non-linearly realized supersymmetry and therefore attention has to be paid when going on-shell. The equation of motion for the auxiliary field gives indeed
\be\label{eq:Fterm}
F = -M^2 - \frac{1}{4M^6}\bar \Omega P_R \Omega \Box (\bar \Omega P_L \Omega) + \frac{3}{16 M^{14}}(\bar\Omega P_R \Omega)(\bar\Omega P_L \Omega) \Box (\bar \Omega P_R \Omega) \Box (\bar \Omega P_L \Omega)\,,
\ee
and the on-shell action is
\be
\begin{aligned}\label{eq:Xaction}
S &=\int d^4x \bigg( - M^4 -\bar \Omega P_L \slashed{\partial}\Omega + \frac{1}{4M^4}\bar \Omega P_L\Omega  \Box (\bar \Omega P_R \Omega)\\
&\quad\qquad\qquad- \frac{1}{16 M^{12}}(\bar\Omega P_R \Omega)(\bar\Omega P_L \Omega) \Box (\bar \Omega P_R \Omega) \Box (\bar \Omega P_L \Omega)\bigg).
\end{aligned}
\ee
By means of a field redefinition between $\lambda$ and $P_L\Omega$, one can prove that this action is equivalent to the Volkov--Akulov model~\cite{Kuzenko:2011tj}.

\subsection{Coupling the Goldstino to gravity}

Superconformal methods are very convenient when constructing supergravity actions. The strategy on which they rely consists in taking advantage of the full superconformal symmetry to fix all of the allowed interactions. This symmetry is then partially broken in order to obtain Poincar\'e supergravity. With such a procedure it is possible to avoid field redefinitions, that might be needed to go to the Einstein frame when using other methods. Hence, in the present work we adopt the superconformal approach to supergravity, following the conventions of~\cite{freedman2012supergravity}.

The superconformal action we are going to consider is of the type
\be
\label{LSUGRA}
S = [-3 X^0 \bar X^0 \e^{-K(X, \bar X)/3}]_D + [(X^0)^3W(X)]_F + [f_{AB}(X)\bar\Lambda^A P_L\Lambda^B]_F,
\ee
 where $\{X^I\}$, $I=0,\dots,n$, is a set of chiral multiplets with $X^0$ the compensator, $\Lambda^A$, $A=1,\dots n_v$, is a set of vector multiplets, $K$ is the K\"ahler potential, $W$ the superpotential and $f_{AB}$ the gauge kinetic function. The compensator has Weyl weight 1, while the other chiral multiplets have Weyl weight 0. In order to obtain Poincar\'e supergravity one has to fix $X^0 = \kappa^{-1} e^{\frac K6}$, which introduces the Planck scale into the theory.

A minimal model in which the Goldstino is coupled to gravity is given by
\be
K =  X \bar X, \qquad W = W_0+M^2 X,
\ee
where $X$ is the nilpotent Goldstino multiplet introduced before. In the case in which there are no vector multiplets, the action~\eqref{LSUGRA} reduces to
\be
\label{dSsugra}
S = [-3X^0 \bar X^0 + X^0\bar X^0 X\bar X]_D + [(X^0)^3(W_0 + M^2 X)]_F, \qquad X^2=0.
\ee
This is the generalization of~\eqref{LVAKS} to local supersymmetry and it has been studied in~\cite{Farakos:2013ih,Dudas:2015eha,Bergshoeff:2015tra,Hasegawa:2015bza,Ferrara:2015gta}.
The model is sometimes called (pure) de Sitter supergravity because the only propagating modes are the graviton and the gravitino, the Goldstino being a pure gauge degree of freedom. In addition, for certain values of the parameters in the scalar potential, the cosmological constant is positive.
We stress again that, when calculating the component form of~\eqref{dSsugra}, it is important to substitute $X = \frac{\bar \Omega P_L\Omega}{2F}$ before going on-shell, since this will contribute to the equations of motion of the auxiliary fields.

\subsection{Other constrained multiplets}\label{sec:othercm}
A general procedure to constrain supersymmetric multiplets and remove any desired component has been given in~\cite{DallAgata:2016yof}. In the following, besides the nilpotent Goldstino multiplet $X$, we are going to use other types of constrained multiplets, which we briefly review here.
Notice that it is possible to implement them dynamically at the Lagrangian level, by means of a Lagrange multiplier~\cite{Ferrara:2016een}. In this way supersymmetry remains linear off-shell.

\subsubsection{Constrained chiral multiplets $Y^i$} Given a set of chiral multiplets $Y^i=\{Y^i, P_L \Omega^i, F^i\}$, by imposing the constraints~\cite{Brignole:1997pe,DallAgata:2015pdd}
\be
\label{orthconstr}
X^2=0,\qquad XY^i=0,
\ee
the scalar fields in the lowest components of $X$ and $Y^i$ are removed and expressed as
\be
X = \frac{\bar\Omega P_L\Omega}{2F},\qquad Y^i = \frac{\bar \Omega^i P_L\Omega}{F}-\frac{\bar \Omega P_L \Omega}{2F^2}F^i.
\ee
These multiplets contain therefore only fermions as propagating degrees of freedom. They have been used in~\cite{Kallosh:2016aep,Vercnocke:2016fbt} to describe the world volume spinors of an anti-D3-brane.

\subsubsection{Constrained chiral multiplets $H^a$}\label{subsec:constrH} Given another set of chiral multiplets $H^a=\{H^a, P_L \Omega^{a}, F^{a}\}$, by imposing the constraints~\cite{Komargodski:2009rz}
\be
X^2=0,\qquad X\bar H^a = \text{chiral},
\ee
the fermion and the auxiliary field in $H^a$ are removed and expressed as
\begin{align}
P_L\Omega^{a} &= \frac{\slashed{\mathcal{D}}H^a}{\bar F}P_R\Omega,\\
F^{a} &= \mathcal{D}_\mu \left(\frac{\bar \Omega}{\bar F}\right) \gamma^\nu \gamma^\mu \left(\frac{P_R \Omega}{\bar F}\right) \, \mathcal{D}_\nu H^a +\frac{\bar \Omega P_R \Omega}{2\bar F^2}\Box H^a.
\end{align}
The chiral multiplets $H^a$ contain therefore only a complex scalar as independent component field. Notice that, due to this fact, a superpotential of the type $W=W(H)$ does not lead to mass terms for the scalars $H^a$, but to fermionic terms containing Goldstino interactions.

\subsubsection{Constrained chiral field strength multiplet $P_L\Lambda_\alpha$} 

The field strength chiral multiplet that has the gaugino as its lowest component is
\be
\label{Lambdamult}
P_L\Lambda_\alpha = \{P_L\Lambda_\alpha, (P_L\chi)_{\beta\alpha}, F^\Lambda_\alpha \},
\ee
where
\begin{align}
(P_L\chi)_{\beta\alpha} &= \sqrt 2\left[-\frac14 (P_L\gamma^{ab}C)_{\beta\alpha}\hat F_{ab} + \frac i2 {\rm D} (P_LC)_{\beta\alpha}\right],\\
F_\alpha^\Lambda &= (\slashed{\mathcal{D}}P_R\Lambda)_\alpha
\end{align}
and where we have explicitly written the spinorial indices to avoid confusion. $C_{\alpha \beta}$ satisfying $C^T = - C$ is the matrix used to raise and lower fermionic indices, while $\hat F_{ab} =  e_a^\mu e_b^\nu (2\partial_{[\mu} A_{\nu]} + \bar \psi_{[\mu }\gamma_{\nu]}\lambda)$ is the covariant vector field strength and D the real auxiliary field. This multiplet is the analogous of the superfield strength $W_\alpha = -\frac14 \bar D^2 D_\alpha V$ defined in superspace, which indeed is chiral and has the gaugino in the lowest component.

The gaugino can be eliminated by imposing the constraint
\be
\label{constrXW}
X P_L \Lambda_\alpha =0,
\ee
which gives
\be
\begin{aligned}
P_L\Lambda_\alpha &= \frac 1F \left(\bar \Omega P_L\chi\right)_\alpha - \frac XF {\slashed{\mathcal{D}}_\alpha}^\beta \left(\frac{(\bar\Omega P_R \chi)_\beta}{\bar F}\right)\\
&+\frac{X}{F}{\slashed{\mathcal{D}}_\alpha}^\beta \left(\frac{\bar X}{\bar F}{\slashed{\mathcal{D}}_\beta}^\gamma \left(\frac{(\bar \Omega P_L \chi)_\gamma}{F}\right)\right)\\
& - \frac{X\bar X}{F^2\bar F^2}{(\slashed{\mathcal{D}}\slashed{\mathcal{D}}X)_\alpha}^\beta {(\gamma^\mu)_\beta}^\delta (\mathcal{D}_\mu \bar \Omega P_R \chi)_\delta,
\end{aligned}
\ee
where $X = \frac{\bar \Omega P_L\Omega}{2F}$ and $(\bar\Omega P_L\chi)_\alpha=\Omega^\beta {(P_L)_\beta}^\gamma\chi_{\gamma\alpha}$. The constrained multiplet $P_L\Lambda_\alpha$ describes therefore only an abelian gauge vector as an independent propagating degrees of freedom. The superspace constraint corresponding to \eqref{constrXW} is $XW_\alpha =0$.

\section{Constructing a supergravity action for the anti-D3-brane}\label{sec:sugraaction}

In this section we recast the anti-D3-brane action~\eqref{SDbcomplete} in the language of $\N=1$ supergravity in four dimensions. The rewriting of the fermionic action coupled to the closed string moduli was done already in~\cite{GarciadelMoral:2017vnz} (see for example their equations (3.51) and (3.52)).\footnote{We thank Flavio Tonioni and the authors of \cite{GarciadelMoral:2017vnz} for alerting us to a problem with their mass term for the fermions. We will rectify this and present below two different ways of writing the fermionic mass term.} Here, we extend this result by considering also the bosonic part, together with the terms that mix world volume bosons and fermions, and the sector containing the U$(1)$ gauge vector. 

The logic consists in embedding each of the world volume fields into one of the constrained multiplets presented in section~\ref{sec:sugracomponent}. In this way we will be able to use the standard language of supergravity, namely to write down a \K potential, a superpotential and a gauge kinetic function, but at the same time the non-linear realization of supersymmetry will be manifest. The very fact that it is possible to use non-linear supersymmetry to rewrite the anti-D3-brane action confirms that the anti-brane is breaking supersymmetry spontaneously.

Note that branes break supersymmetry generically at the string scale. In our case, the anti-D3-brane sits at the bottom of a warped throat and therefore the string scale is warped down compared to the bulk string scale. A recent discussion of these scales can be found for example in \cite{deAlwis:2016cty}. The warped down string scale, in our conventions with $l_s=2\pi\sqrt{\alpha'}=1$, is given by the first term in \eqref{SDbcomplete}: $M_s^4 =2  e^{4 \A-8 u}$. This sets the supersymmetry breaking scale, as can be seen by looking for example at equations \eqref{eq:Fterm} and \eqref{eq:Xaction} above. One expects that at this scale linear supersymmetry will be restored and indeed massive open string states arise as new degrees of freedom. The particular Klebanov-Strassler throat geometry~\cite{Klebanov:2000hb}, which has been intensively studied in the KKLT context, has a three-sphere at the bottom of the warped throat. Anti-D3-branes at the bottom of the throat can then decay via nucleation of an NS5-brane that is wrapping an $S^2$ inside the $S^3$ \cite{Kachru:2002gs}. Such a decay leads to a supersymmetric state and one can actually write down a supergravity theory with linear supersymmetry by including the infinite tower of Kaluza–Klein (KK) modes associated with the $S^3$ \cite{Aalsma:2018pll}. Therefore, in this particular case one finds that new states come in already below the supersymmetry breaking scale and lead to a restoration of linear supersymmetry. \\Having the supersymmetry breaking scale at the warped down string scale, which is above the warped down KK scale at which the four-dimensional effective field theory breaks down, might seem worrisome. However, the hallmark of a supergravity theory is the presence of a gravitino and the mass of the gravitino in the KKLT scenario can be well below the KK scale. Thus, a description in terms of a four dimensional $\N=1$ theory is appropriate.

\subsection{Goldstino and matter component fields}

We start by considering the couplings involving scalars and fermions, while we will focus on the gauge vector in section \ref{subsec:vecsugra}. In particular, we generate the mass terms for the fermions with a different mechanism with respect to \cite{GarciadelMoral:2017vnz} and we show how to also include the anti-D3-brane world volume scalars $H^a$ in the supergravity action. An alternative way for producing a fermionic mass term is presented in section \ref{subsec:Mhat}.

The first step is to embed the Goldstino $\lambda$ and the triplet of fermions $\chi^i$ into, respectively, a chiral multiplet $X$ and a triplet of chiral multiplets $Y^i$ satisfying the constraints~\eqref{orthconstr}. The kinetic terms of the spin-1/2 fields in \eqref{SDbcomplete} can then be generated from the following K\"ahler potential, in which the bulk moduli are coupled to world volume fermions \cite{GarciadelMoral:2017vnz}
\begin{equation}
\begin{aligned}\label{GMPQZ}
K=&-\log(-{\rm i}(\tau-\bar{\tau}))-3\log\left[ (-\rmi (T-\bar T))f(U^A,\bar{U}^A)^\frac13\right]\\
&-3\log\left(1-\frac{e^{-4{\cal A}}X\bar{X}}{3(-{\rm i}(\tau-\bar{\tau}))(-\rmi(T-\bar T))f(U^A,\bar U^A)}\right.\\
&\left.\qquad\qquad-\frac{e^{-4{\cal A}}\delta_{i\bar\jmath}Y^i\bar{Y}^{\bar \jmath}}{3(-{\rm i}(\tau-\bar{\tau}))(-\rmi (T-\bar T))^2f(U^A,\bar{U}^A)^\frac13}\right).
\end{aligned}
\end{equation}
We use from now on $f(U^A, \bar U^A)=-\rmi \int \Omega \w \bar \Omega\,$ to avoid confusion between the holomorphic (3,0)-form and the fermions $\Omega$, $\Omega^i$. The couplings of $X$ and $Y^i$ to the bulk moduli are fixed as follows. The coupling to $\tau$ is determined by requiring modular invariance for the world volume action (see subsection \ref{subsec:modinv} below for details). For what concerns the other moduli, the couplings to $X$ are fixed by matching with the scalar potential in~\eqref{SDbcomplete}, while those to $Y^i$ are fixed by matching with the \K covariant kinetic terms of the massive spin-1/2 world volume spinors (cf. subsection \ref{sec:D3K}). In particular, the fermions $\Omega^i$ inside $Y^i$ are related to $\chi^i$ by the field redefinitions 
\begin{align}
\label{fermionredef}
P_L\Omega^i &= {2}{\rm i} e^{4\A-\frac{\phi}{2}}  f(U^A,\bar{U}^A)^{\frac16}P_L \chi^i + \dots,
\end{align}
where dots stand for higher order terms. Notice that we are not matching the supergravity expression with the kinetic term of $\lambda$ in \eqref{SDbcomplete} since, due to the fact that such a fermion is a Goldstino, its couplings are not physical and they can be set to zero by going to the unitary gauge. For what follows, it is sufficient to keep in mind that, in our supergravity description, the Goldstino resides in the multiplet $X$, namely $P_L \Omega \sim P_L \lambda +\dots$, where dots stand for higher order terms. We stress that the presence of the Goldstino is an essential feature of the anti-D3-brane and a similar reasoning cannot be repeated in the case of the D3-brane.

The superpotential that sources the supersymmetry breaking and gives rise to the anti-D3-brane uplift term is
\be\label{WSUGRA}
W =  W_{GVW} + W_{np} + M^2 X,
\ee
where $M^2 = \sqrt{2}$. The parameter $M$ is related to the SUSY breaking scale, which is the warped down string scale. By rescaling $X$ we have included the warp factor in the \K potential and we have set the anti-D3-brane tension to $T_{D3} = 2\pi = M^4\pi$. The very form of the superpotential \eqref{WSUGRA} implies that supersymmetry is spontaneously broken by the auxiliary field of $X$ and therefore it is consistent to identify the Goldstino $\lambda$ with the fermion $\Omega$ inside $X$, at leading order.

The \K potential and superpotential presented so far reproduce correctly the kinetic terms for the fermions and the scalar potential in \eqref{SDbcomplete}. On the other hand, at this stage the fermions are massless in the supergravity theory, since none of the couplings we introduced is producing a mass term for them. In order to give a mass to the fermion triplet, it was proposed in~\cite{Vercnocke:2016fbt, GarciadelMoral:2017vnz} to add a contribution $W_{m}=h_{ij}Y^iY^j$ to the superpotential. However, when the axio-dilaton is dynamical and not integrated out, then such a mass term would require $h_{ij}\propto \bar G_3$ to be anti-holomorphic in $\tau$, which seems incompatible with supersymmetry. Instead of adding a term to the superpotential, one can therefore follow a different strategy and modify the \K potential. The required modification of $K$ given in equation \eqref{GMPQZ} and that generates the desired fermionic mass term is
\begin{align}
\label{GMPQZ2}
K=&-\log(-{\rm i}(\tau-\bar{\tau}))-3\log\left[ (-\rmi (T-\bar T))f(U^A,\bar{U}^A)^\frac13\right]\nn\\
&-3\log\left(1-\frac{e^{-4{\cal A}}X\bar{X}}{3(-{\rm i}(\tau-\bar{\tau}))(-\rmi(T-\bar T))f(U^A,\bar U^A)}\right.\nn\\
&\left.\qquad\qquad -\frac{e^{-4{\cal A}}\delta_{i\bar\jmath}Y^i\bar{Y}^{\bar \jmath}}{3(-{\rm i}(\tau-\bar{\tau}))(-\rmi (T-\bar T))^2f(U^A,\bar{U}^A)^\frac13}\right.\\
&\qquad\qquad\left. +\frac{e^{-8{\cal A}}\lp m_{ij}\bar X Y^iY^j+\overline{m}_{\ib\jb} X \bar Y^\ib  \bar Y^\jb\rp}{6M^2 (-\rmi (\tau-\bar \tau))^{\frac32}(-\rmi(T-\bar T))^\frac32f(U^A,\bar U^A)^{\frac56}}\right),\nn
\end{align}
where $m_{ij}$ is given in equation \eqref{mij} above. As we discuss more extensively in section \ref{subsec:modinv}, this modification of the \K potential does not spoil the modular invariance of the supergravity action.
In section \ref{subsec:Mhat} we presented an alternative mechanism, in which the mass to the fermion triplet is given by a superpotential term. However, such a construction requires some technical explanation in order to be presented properly and this is the reason why we postpone it for the time being.

We present now how to introduce the dependence on the world volume scalars $H^a$ parameterizing small fluctuations around the position of the anti-D3-brane in the warped throat. The kinetic term of $H^a$ arises from the DBI action, which is the same for D3-branes and anti-D3-branes. It is known to be well described by the \K potential in~\eqref{eq:Kpot} in the case of fixed complex structure moduli and we have shown that it should be modified as in equation \eqref{eq:D3Kahler} when the complex structure and axio-dilaton are dynamical. We can now consider a simple prescription to generalize the supergravity \K potential~\eqref{GMPQZ2}. We first embed the scalar fields into the constrained multiplets $H^a$ introduced in subsection~\ref{subsec:constrH}, that contain only a scalar in the lowest component as an independent degree of freedom. Then we let $\A$ depend generically on $H^a$ and we formally shift the volume modulus as  $-{\rm i}(T-\bar{T})\to -{\rm i}(T-\bar{T})+k(H^a,\bar{H}^a)/f(U^A,\bar U^A)^{\frac13}=e^{4u}$ to match the results. As a result, we get the following \K potential of $\N=1$ supergravity in four dimensions
\be
\begin{aligned}
K=&-\log(-{\rm i}(\tau-\bar{\tau}))-3\log\left[ (-\rmi (T-\bar T))f(U^A,\bar{U}^A)^\frac13+k(H^a,\bar H^a)\right]\\
&-3\log\left(1-\frac{e^{-4{\cal A}(H^a,\bar H^a)-4u}}{3(-{\rm i}(\tau-\bar{\tau}))f(U^A,\bar U^A)}X\bar{X}
-\frac{e^{-4{\cal A}(H^a,\bar H^a)-8u}}{3(-{\rm i}(\tau-\bar{\tau}))f(U^A,\bar U^A)^\frac13}\delta_{i\bar\jmath}Y^i\bar{Y}^{\bar \jmath}\right.\\
&\qquad\qquad\left. +\frac{e^{-8{\cal A}(H^a,\bar H^a)-6u}\lp m_{ij}\bar X Y^iY^j+\overline{m}_{\ib\jb} X \bar Y^\ib  \bar Y^\jb\rp}{6M^2 (-\rmi (\tau-\bar \tau))^{\frac32}f(U^A,\bar U^A)^{\frac56}}\right).
\end{aligned}
\ee
As a consequence of this last step, the superpotential $W_{np}$ is also getting a dependence on the scalars $H^a$. However, as we explained previously, the net result of this additional dependence would be to produce corrections which are highly suppressed in the regime we are considering. For all of the purposes of the present work, we can therefore neglect the dependence of $W_{np}$ on $H^a$ and keep considering~\eqref{WSUGRA} as the expression for the superpotential.

Having identified the \K potential and the superpotential, we can use the rules of $\mathcal{N}=1$ supergravity to calculate the scalar potential. The result is of the form
\be
V = V_{KKLT} + V_{\Db},
\ee
where $V_{KKLT}$ contains the contributions from the supergravity bulk fields, while $V_{\Db}$ is the uplift term coming from the anti-D3-brane
\begin{equation}
V_{\Db}=\frac{M^4e^{4{\cal A}(H^a,\bar H^a)}}{\left(-{\rm i}(T-\bar{T})+k(H^a,\bar H^a)f(U^A,\bar U^A)^{-\frac13}\right)^2}=2 e^{4{\cal A}(H^a,\bar H^a)-8u},
\end{equation}
which reproduces~\eqref{eq:scalarpot}, as desired. We recall that to calculate the scalar potential with the constrained multiplets $X$ and $Y^i$, it is sufficient to perform the calculation in the usual manner and then set $X= Y^i=0$ in the final result.

\subsection{Gauge vector field and theta term}\label{subsec:vecsugra}

We focus now on the part of the action containing the world volume gauge vector field. As we mentioned in section  \ref{sec:bosonic}, the kinetic term originating from the DBI action is the same as that of the D3-brane vector, whereas the CS term has the opposite sign due to the difference in the RR charge. As a consequence of this sign flip, it seems that the gauge kinetic function needs to depend on an anti-chiral multiplet, $\bar f=\bar f(\bar{\tau})$, and this would be an obstruction to rephrasing the anti-brane vector field into an $\N=1$ supersymmetric language. However, as we are going to show, the fact that supersymmetry is spontaneously broken and non-linearly realized allows us to also embed correctly the CS term. In this subsection, therefore, we show explicitly how to describe such a CS term with the appropriate sign for the anti-D3-brane case.

As a first step, we embed the world volume vector field into a chiral field strength multiplet $P_L \Lambda_\alpha$ and we constrain it as in~\eqref{constrXW}. 
As explained before, this constraint is removing the gaugino, leaving only the U$(1)$ vector as independent physical degrees of freedom. If we consider then the standard supergravity action for the vector multiplet \cite{freedman2012supergravity}\footnote{In \cite{freedman2012supergravity} the overall factor $-\frac14$ is understood, but we prefer to keep it explicit for convenience.}
\be
\begin{aligned}
\label{LV}
-\frac14 [f(\tau) \bar \Lambda P_L \Lambda]_F &= \int d^4x\, \sqrt{-g_4}\left(-\frac{{\rm Re}(f)}{4} F_{\mu\nu}F^{\mu\nu} + \frac{{\rm Im}(f)}{8}\frac{\epsilon^{\mu\nu\rho\sigma}}{\sqrt{-g_4}} F_{\mu\nu}F_{\rho\sigma}\right.\\
&\qquad \qquad \qquad \qquad \left.+\frac{{\rm Re}(f)}{2}\,{\rm D}^2+\dots\right), 
\end{aligned}
\ee
with $f(\tau) = -{\rm i}\tau$ and where the dots stand for fermionic terms, we notice that the term proportional to ${\rm Im}(f) = - {\rm Re}(\tau)$ has the opposite sign compared to~\eqref{SDbcomplete}. We would therefore like to flip this sign by subtracting from the action twice the same contribution. In order to perform this step in a supersymmetric way, we construct a deformation of the action which is similar in spirit to the recently proposed new Fayet--Iliopoulos D-terms in supergravity~\cite{Cribiori:2017laj,Kuzenko:2018jlz,Antoniadis:2018cpq,Antoniadis:2018oeh,Aldabergenov:2018nzd,Cribiori:2018dlc,Aldabergenov:2019hvl,Kuzenko:2019vaw}. In~\cite{Cribiori:2017laj} it has been shown how to deform~\eqref{LV} and introduce a coupling linear in the auxiliary field D, without spoiling the gauge invariance and without requiring the gauging of the R-symmetry, which is needed for the standard Fayet--Iliopoulos D-term. This new coupling shifts the vacuum expectation value of D and, as a consequence, supersymmetry is spontaneously broken by a D-term. In the same spirit, we would like to introduce a coupling $-\frac14 {\rm Im}(f) \epsilon^{\mu\nu\rho\sigma}\, F_{\mu\nu}F_{\rho\sigma}$, together with all of the additional interactions required by supersymmetry, in order to flip the sign in the theta term in \eqref{LV}. In the following we directly give the expression of such a deformation, but more details about its derivation and on new D-terms in general are given in appendix~\ref{appB:newDterm}.

Given a constrained field strength multiplet $P_L\Lambda_\alpha$, in order to reproduce the vector multiplet interactions of~\eqref{SDbcomplete}, we propose the following supergravity action 
\begin{equation}
\label{eq:vector}
\begin{aligned}
S_{V} =& -\frac14 [f(\tau)\bar \Lambda P_L \Lambda]_F\\
&+\bigg[\frac{X \bar X \left(X^0 \bar X^0 e^{-\frac K3}\right)^3}{\Sigma (\bar X^0e^{-\frac K6}\bar X) \bar \Sigma(X^0 e^{-\frac K6}X)}\, {\rm Im}(f)\,{\rm Im}\left(\Sigma(\bar \omega^2)\right)\bigg]_D, 
\end{aligned}
\end{equation}
where we defined the multiplets 
\be
\omega^2 = \frac{\bar \Lambda P_L \Lambda}{\left(X^0\bar X^0e^{-\frac K3}\right)^2}, \qquad \bar\omega^2 = \frac{\bar \Lambda P_R \Lambda}{\left(X^0\bar X^0e^{-\frac K3}\right)^2},
\ee
where ($\bar \Sigma$)$\Sigma$ is the (anti-)chiral projector in the superconformal setup. We always assume X to be nilpotent. This action is made up of the standard kinetic coupling for the vector multiplet \eqref{LV} and by a second, novel term, whose origin is presented in appendix~\ref{appB:newDterm}. The property of this coupling that is important for the present discussion is that its component expansion starts precisely with the desired theta term
\be
\begin{aligned}
\label{tterm}
S_{\theta-\text{term}}&=\left[\frac{X \bar X \left(X^0 \bar X^0 e^{-\frac K3}\right)^3}{\Sigma (\bar X^0e^{-\frac K6}\bar X) \bar \Sigma(X^0 e^{-\frac K6}X)}\, {\rm Im}(f)\,{\rm Im}\left(\Sigma(\bar\omega^2)\right)\right]_D \\
&= -\frac14 \int d^4x\, \, {\rm Im}(f) \epsilon^{\mu\nu\rho\sigma} F_{\mu\nu}F_{\rho\sigma}+\dots
\end{aligned}
\ee
and the dots stand for fermionic terms. After fixing the superconformal symmetry with $X^0 = \kappa^{-1}e^{\frac K6}$, the bosonic sector of \eqref{eq:vector} reduces to the desired result ($\kappa=1$)
\begin{equation}
S_{V,\,bos}= \int d^4x \sqrt{-g_4}\left[-\frac{1}{4}{\rm Im}(\tau) F^{\mu\nu}F_{\mu\nu}+\frac{1}{8}{\rm Re}(\tau)\frac{\epsilon^{\mu\nu\rho\sigma}}{\sqrt{-g_4}}F_{\mu\nu} F_{\rho\sigma}\right],
\end{equation}
where we used $f(\tau) = -{\rm i} \,\tau$ and we integrated out the auxiliary field D.
In particular, the sign of the theta term has been flipped since the contribution coming from the second term in~\eqref{eq:vector} is minus twice that arising from the first term.

A few comments are in order at this point. We find that \eqref{eq:vector} correctly realizes the bosonic part of the world volume vector action~\eqref{eq:bosonic}. It is worth noting that the second term in~\eqref{eq:vector}, which is essential in order to realize the correct CS term, can be consistently introduced only if the auxiliary field $F$ of $X$ is non-vanishing. This condition is always satisfied within our setup, since the anti-D3-brane is breaking supersymmetry spontaneously and the fermion in $X$ provides the Goldstino. Notice also that the coupling~\eqref{tterm} contains terms quadratic (and also of higher order) in the fermions in its component expansion, which might jeopardize the matching of the anti-D3-brane action with our supergravity proposal. However, due to the non-linear realization of supersymmetry, the fermionic terms will be functions of the Goldstino and vanish identically in the unitary gauge. Therefore, even if differences might be present in the fermionic couplings between the anti-D3-brane action and the supergravity one, these differences are not physical and can be removed by an appropriate gauge choice. We stress that this is true as long as the fermion in $X$ is completely aligned with the Goldstino.

By exploiting the properties of the operator $\Sigma$ and the constraints \eqref{X2=0} and \eqref{constrXW}, it is possible to recast the action \eqref{eq:vector} into a more suggestive form. Notice first that 
\begin{equation}
[f(\tau)\bar \Lambda P_L \Lambda]_F = \left[\frac{\Sigma \left(\bar X^0 e^{-\frac K6}\bar X f(\tau)\right)}{\Sigma \left(\bar X^0 e^{-\frac K6}\bar X \right)}\bar \Lambda P_L \Lambda \right]_F =\left[\Sigma \left( \frac{\bar X^0 e^{-\frac K6}\bar X f(\tau)}{\Sigma \left(\bar X^0 e^{-\frac K6}\bar X \right)}\right)\bar \Lambda P_L \Lambda \right]_F,
\end{equation}
where we used the fact that $\Sigma(AB) = A \Sigma(B)$, if $A$ is chiral and B has weights $(w,w-2)$, as proved in \cite{Ferrara:2016een}.
Then, since $P_L \Lambda_\alpha$ is constrained, we have
\be
\begin{aligned}
S_{\theta-\text{term}}&=\left[\frac{X \bar X \left(X^0 \bar X^0 e^{-\frac K3}\right)^3}{\Sigma (\bar X^0e^{-\frac K6}\bar X) \bar \Sigma(X^0 e^{-\frac K6}X)}\, {\rm Im}(f)\,{\rm Im}\left(\Sigma(\bar\omega^2)\right)\right]_D \\
&=\left[\frac{\rmi}{2} \frac{\bar X^0 e^{-\frac K6} \bar X}{\bar \Sigma(X^0 e^{-\frac K6}X)}{\rm Im}(f) \,\bar \Lambda P_L \Lambda + c.c. \right]_D\\
&=\left[\frac{\rmi}{2}\Sigma \left(\frac{\bar X^0 e^{-\frac K6} \bar X}{\bar \Sigma(X^0 e^{-\frac K6}X)}{\rm Im}(f)\right)\bar \Lambda P_L \Lambda\right]_F,
\end{aligned}
\ee
where in going from the second to the third line we used that $[C]_D = \frac12 [\Sigma(C)]_F$, another property stated for example in \cite{Ferrara:2016een}. Using these results, the two terms in \eqref{eq:vector} can be put together and the vector multiplet action acquires the more familiar form
\be
\begin{aligned}
S_V&=-\frac14 \left[\Sigma \left( \frac{\bar X^0 e^{-\frac K6}\bar X \bar f(\bar \tau)}{\Sigma \left(\bar X^0 e^{-\frac K6}\bar X \right)}\right)\bar \Lambda P_L \Lambda \right]_F\\
&\equiv -\frac14 \left[\hat f_{\Db}(\bar \tau, \bar X) \,\bar \Lambda P_L \Lambda\right]_F,
\end{aligned}
\ee
with $\bar f = {\rm Re}(f)-\rmi {\rm Im}(f)$ and where we defined the composite (or generalized) anti-D3-brane gauge kinetic function 
\be
\label{fD3b}
\hat f_{\Db} = \Sigma \left( \frac{\left(\bar X^0 e^{-\frac K6}\bar X\right) \bar f(\bar \tau)}{\Sigma \left(\bar X^0 e^{-\frac K6}\bar X \right)}\right).
\ee
This is a chiral multiplet that is anti-holomorphic in $\tau$. It is important to notice that $\hat f_{\Db}$ contains Goldstino interactions, which implement the non-linear realization of supersymmetry and are essential in order to consistently couple the vector multiplet to $\bar \tau$. The lowest component of  $\hat f_{\Db}$ is given by
\begin{equation}
\hat f_{\Db} = \bar f(\bar \tau) + \text{fermions}
\end{equation}
and the fermionic terms vanish identically in the unitary gauge in which the Goldstino is set to zero. Indeed, one can equally think of $\hat f_{\Db}$ as a chiral multiplet that satisfies the additional constraint
\be
X \bar X \hat f_{\Db} = X \bar X \bar f(\bar \tau).
\ee

\subsection{Superpotential mass for the triplet of fermions }\label{subsec:Mhat}

In the previous section we have seen how non-linear supersymmetry allows us to consistently couple $\bar f (\bar \tau)= \rm i\bar \tau$ to the vector field, using a manifestly supersymmetric language. This strategy is indeed general and can also be employed to describe the mass term of the fermion triplet. We recall in fact that the obstruction to producing such a term from the superpotential was that, in the case of the anti-D3-brane, it depends on $\bar \tau$ instead of $\tau$.

Inspired by the composite anti-D3-brane gauge kinetic function $\hat f_{\Db}$ introduced before, we define the chiral multiplet
\be\label{eq:MIJ}
\hat M_{ij} =\Sigma \left(\frac{\left(\bar X^0 e^{-\frac K6}\bar X\right)e^{-4\A-2u} \,m_{ij}}{\,\Sigma\left(\bar X^0 e^{-\frac K6}\bar X\right)(-\rmi(\tau-\bar\tau))^\frac12 f(U^A,\bar U^A)^{-\frac16}}\right),
\ee
where $m_{ij}$ is the fermion mass given in \eqref{mij}. The lowest component of this multiplet is
\be
\hat M_{ij} =\frac{e^{-4\A-2u}}{(-\rmi(\tau-\bar\tau))^\frac12 f(U^A,\bar U^A)^{-\frac16}} m_{ij} + \text{fermions}
\ee
and the fermionic terms vanish in the unitary gauge. Indeed, $\hat M_{ij}$ satisfies the constraint
\begin{equation}
 X \bar X \hat M_{ij} = X \bar X \left(\frac{e^{-4\A-2u}m_{ij}}{(-\rmi(\tau-\bar\tau))^{\frac12}f(U^A,\bar U^A)^{-\frac16}}\right).
\end{equation}
A mass for the fermions in $Y^i$ can then be produced by a superpotential holomorphic in $\hat M_{ij}$ and $Y^i$ of the type
\be
W_m (\hat M, Y) = \frac12 \hat M_{ij} Y^i Y^j.
\ee
We stress that this type of constructions, namely $\hat f_{\Db}$ and $\hat M_{ij}$, can be consistently defined only in a setup in which supersymmetry is spontaneously broken and the auxiliary field of $X$ acquires a non-vanishing vacuum expectation value.\footnote{In the language of flat superspace, \eqref{fD3b} and \eqref{eq:MIJ} are defined as the chiral superfields
\be
\hat f_{\Db} = \bar D^2\left(\frac{\bar X \bar f}{\bar D^2 \bar X}\right), \qquad \hat M_{ij} = \bar D^2\left(\frac{\bar X \, e^{-4\A-2u}\,m_{ij}}{\bar D^2 \bar X\, (-\rmi (\tau-\bar\tau))^\frac12 f(U^A,\bar U^A)^{-\frac16}}\right),
\ee
and one has to require $\langle D^2 X\rangle \neq 0$.}

\section{Summary: The supergravity action and its modular invariance} 
\label{sec:summary}

In this section we summarize our results and present some consistency checks on the expected modular properties of the anti-D3-brane action.
We have showed that, up to quadratic terms in the fermions, the anti-D3-brane action \eqref{SDbcomplete} can be described in $\N=1$ supergravity in four dimensions by
\begin{equation}
\label{Ssugracomplete}
\begin{aligned}
S &= [\hat f_{\Db}(\bar \tau, \bar X)\bar \Lambda P_L \Lambda]_F+[-3 X^0 \bar X^0 e^{-\frac K3}]_D +[(X^0)^3W]_F.
\end{aligned}
\end{equation}
The anti-D3-brane gauge kinetic function is built out of the Goldstino and the axio-dilaton and is defined as the chiral multiplet
\be
\hat f_{\Db} = \Sigma \left( \frac{\bar X^0 e^{-\frac K6}\bar X \bar f(\bar \tau)}{\Sigma \left(\bar X^0 e^{-\frac K6}\bar X \right)}\right),
\ee
with $\bar f(\bar \tau) = \rmi \bar \tau$. We presented two different expressions for the \K potential and the superpotential, depending on how the mass term for the fermions in $Y^i$ is generated. They lead to the same physical action and are related by field redefinitions involving the Goldstino. 
One possibility is to generate the mass for the triplet of fermions from the \K potential
\be
\label{eq:Kfinal1}
\begin{aligned}
K_1=&-\log(-{\rm i}(\tau-\bar{\tau}))-3\log\left[ (-\rmi (T-\bar T))f(U^A,\bar{U}^A)^\frac13+k(H^a,\bar H^a)\right]\\
&-3\log\left(1-a X \bar X - b \delta_{i\jb} Y^i \bar Y^\jb + c\lp m_{ij}\bar X Y^iY^j+\overline{m}_{\ib\jb} X \bar Y^\ib  \bar Y^\jb\rp\right),
\end{aligned}
\ee
where $f(U^A, \bar U^A)=-\rmi \int \Omega \w \bar \Omega\,$ and 
\begin{align}
a &= \frac{e^{-4{\cal A}(H^a,\bar H^a)}}{3(-{\rm i}(\tau-\bar{\tau}))(-\rmi(T-\bar T)+k(H^a,\bar H^a)f(U^A,\bar U^A)^{-\frac13})f(U^A,\bar U^A)},\\
b &= \frac{e^{-4{\cal A}(H^a,\bar H^a)}}{3(-{\rm i}(\tau-\bar{\tau}))(-\rmi (T-\bar T)+k(H^a,\bar H^a)f(U^A,\bar{U}^A)^{-\frac13})^2f(U^A,\bar{U}^A)^\frac13},\\
c& = \frac{e^{-8{\cal A}(H^a,\bar H^a)}}{6M^2 (-\rmi (\tau-\bar \tau))^{\frac32}(-\rmi(T-\bar T)+k(H^a,\bar H^a)f(U^A,\bar U^A)^{-\frac13})^{\frac32}f(U^A,\bar U^A)^{\frac56}}.
\end{align}
In this case we then choose the superpotential to be
\be
\label{eq:Wfinal1}
W_1 =  W_{GVW} + W_{np} + M^2 X.
\ee
As an alternative, one can take a \K potential of the type
\be
\label{eq:Kfinal2}
\begin{aligned}
K_2=&-\log(-{\rm i}(\tau-\bar{\tau}))-3\log\left[ (-\rmi (T-\bar T))f(U^A,\bar{U}^A)^\frac13+k(H^a,\bar H^a)\right]\\
&-3\log\left(1-a X \bar X - b \delta_{i\jb} Y^i \bar Y^\jb\right),
\end{aligned}
\ee
and give a mass to the fermions using the superpotential
\be
\label{eq:Wfinal2}
W_2 =  W_{GVW} + W_{np} + M^2 X + \frac12\hat M_{ij} Y^i Y^j,
\ee
where 
\be
\hat M_{ij} = \Sigma \left(\frac{\left(\bar X^0 e^{-\frac K6}\bar X\right)e^{-4\A-2u} \,m_{ij}}{\,\Sigma\left(\bar X^0 e^{-\frac K6}\bar X\right)(-\rmi(\tau-\bar\tau))^\frac12 f(U^A,\bar U^A)^{-\frac16}}\right).
\ee
In both cases $M^2 = \sqrt{2}$ and $m_{ij}$ is given in \eqref{mij}. The supergravity multiplets $X$, $Y^i$, $P_L\Lambda_\alpha$ and $H^a$ describing the world volume fields are constrained as
\be
X^2=0,\qquad XY^i=0,\qquad XP_L\Lambda_\alpha=0,\qquad X\bar H^a=\text{chiral}
\ee
and contain respectively, the Goldstino, the triplet of massive fermions, the U$(1)$ gauge vector and the three complex scalars as independent physical degrees of freedom.

Given a \K potential and a superpotential, one can use the standard rules of $\N=1$ supergravity to calculate the scalar potential ${V}^{\rm SUGRA} = e^K\left(|DW|^2-3|W|^2\right)$. This will contain an uplift term which is matching exactly with the anti-D3-brane contribution \eqref{eq:scalarpot}
\be
{V}^{\rm SUGRA} \supset V_{\Db}(H, \bar H) =2  e^{4 \A(H,\bar H)-8 u}.
\ee

\subsection{On the modular invariance of the Goldstino and matter sector}
\label{subsec:modinv}

As a consistency check for our result, we would like to analyze the behavior of the action \eqref{Ssugracomplete} under modular transformations. Since the analysis of the $X$, $Y^i$ sector is quite different from that of the field strength multiplet $P_L\Lambda_\alpha$, we start from the former and we discuss the latter separately afterwards. 

Therefore, neglecting for the moment the couplings involving the vector field, the original anti-D3-brane action \eqref{SDbcomplete} has to be invariant under SL$(2,\mathbb{R})$ transformations
\begin{equation}
\label{modtransf}
\tau \to \frac{a\tau+b}{c\tau +d}, \qquad G_3 \to \frac{G_3}{c\tau+d}.
\end{equation}
These imply that the world volume fermions have to transform as
\begin{equation}
\label{fermtransf}
P_L\lambda \to e^{-\rmi \delta } P_L\lambda, \qquad P_L\chi^i \to e^{-\rmi \delta} P_L \chi^i,
\end{equation}
where we defined the phase \cite{GarciadelMoral:2017vnz}
\begin{equation}
e^{-2\rmi \delta} = \left(\frac{c\bar \tau +d}{c\tau + d}\right)^\frac12.
\end{equation}
In particular, the transformation of the triplet of fermions $P_L\chi^i$ can be deduced by looking for example at their mass term in \eqref{SDbcomplete}. Indeed, \eqref{fermtransf} together with 
\be
m_{ij}  \to e^{2i\delta} m_{ij},
\ee
which follows from \eqref{modtransf}, imply that $m_{ij} \bar \chi^i P_L\chi^j$ is modular invariant. Once the transformation of $P_L\chi^i$ is fixed, it seems natural to let $P_L\lambda$ transform in the same way, since these four fermions resided originally in the same $\N=4$ multiplet. However, as we stressed previously, it is important to keep in mind that the Goldstino component field is not a physical degree of freedom in supergravity. Therefore we will not insist on studying its modular properties further at the component level; instead we will gain information by analyzing the couplings of the multiplet $X$, which beside the Goldstino contains also the auxiliary field $F$.

Since $G_3$ is transforming under modular transformations, for consistency of the superpotential of the supergravity theory we have to require that the Goldstino multiplet transforms as well, namely
\begin{equation}
\label{Xtransf}
X \to \frac{X}{c\tau +d}.
\end{equation}
Such a requirement, in turn, implies that $\frac{X\bar X}{\tau-\bar \tau}$ is modular invariant, since $
{\rm Im} (\tau) \to \frac{{\rm Im}(\tau)}{|c\tau+d|^2}$, and it fixes  the coupling between the axio-dilaton and the nilpotent multiplet $X$ in the \K potential according to the following reasoning. For a variation of the superpotential of the type
\begin{equation}
W \to \frac{W}{c\tau +d},
\end{equation}
the \K potential is allowed to vary as 
\begin{equation}
K \to K + \log|c\tau+d|^2,
\end{equation}
in order that, if the supergravity theory is \K invariant, then it is also modular invariant.
From the form of the \K potential \eqref{eq:Kfinal1} we can see that all of the freedom of $K$ to transform under modular transformations is already exhausted by the first term, namely $-\log(-\rmi(\tau-\bar \tau))$, and thus all of the remaining couplings have to be modular invariant. With a similar argument, we can conclude that $\delta_{i\bar\jmath}\frac{Y^i \bar Y^{\bar \jmath}}{\tau-\bar\tau}$ is modular invariant if $Y^i$ transforms as $X$, namely
\begin{equation}
\label{Ytransf}
Y^i \to \frac{Y^i}{c\tau+d}.
\end{equation}
By direct inspection one can finally check that all of the remaining couplings in the \K potential and in the superpotential have the correct behaviour under modular transformations, for both the choices of $K$ and $W$ that we presented before. In particular, in the case in which the multiplet $X$ is nilpotent, it is iimmediatly seen that $\frac{X}{\bar \Sigma(X^0 X)}$ is modular invariant, indeed
\be
\frac{X}{\bar \Sigma(X^0 X)} \to \frac{X/(c\tau+d)}{\bar \Sigma(X^0 X/(c\tau+d))} = \frac{X}{\bar \Sigma(X^0 X)},
\ee
where we used the fact that the terms in which the operator $\bar \Sigma$ acts on $\tau$ are vanishing due to the constraint $X^2=0$. 
This observation can be helpful in proving the modular invariance of the proposed supergravity action, in particular for what concerns the superpotential mass term for the fermions.

\subsection{Self-duality of the vector}

As explained in the seminal paper \cite{Gaillard:1981rj}, when a vector field is coupled to the axio-dilaton, the original U$(1)$ duality group is enhanced to SL$(2,\mathbb{R})$. This corresponds precisely to the group of modular transformations we discussed so far. With respect to the previous discussion, however, under duality rotations the action of the vector multiplet is not expected to be invariant. Indeed, the authors of \cite{Gaillard:1981rj} calculated the general form of the induced variation. A duality transformation is in fact a symmetry of the equations of motion of the vector field, which is sometimes called self-duality since it exchanges the electric field strength with its magnetic dual and the gauge coupling with its inverse. In the following we check that the vector multiplet part of the anti-D3-brane action enjoys this property, namely that it is on-shell equivalent to an action of the same functional form, in which the vector field is exchanged with its dual and the gauge kinetic function with its inverse: 
\be
P_L\Lambda_\alpha \longleftrightarrow P_L\Lambda_{D\,\alpha},\qquad\qquad \hat f_{\Db} \longleftrightarrow (\hat f_{\Db})^{-1}.
\ee

Since self-duality is an on-shell property, we are allowed to use any form of the action which reduces on-shell to the vector multiplet part of the anti-D3-brane action
\be
\label{SV}
S_V = -\frac14 \left[\hat f_{\Db} \bar\Lambda P_L \Lambda\right]_F.
\ee
In particular, it is convenient to relax the constraint \eqref{constrXW} on the vector multiplet and impose it by means of a Lagrange multiplier. We consider therefore the action
\be
\label{tildeSV}
\tilde S_V = -\frac14 \left[\hat f_{\Db} \bar\Lambda P_L \Lambda\right]_F + \frac12[\bar \Phi P_L \Lambda X]_F +\left[\frac i2 \bar \Lambda_D P_L \Lambda\right]_F,
\ee
where $P_L \Lambda_\alpha$ is chiral but otherwise unconstrained, $P_L\Phi_\alpha$ is a Lagrange multiplier chiral multiplet implementing the constraint \eqref{constrXW} and $P_L\Lambda_{D\,\alpha}$ is the dual of $P_L \Lambda_\alpha$. In particular, the chiral multiplet $P_L\Lambda_\alpha$ does not satisfy any Bianchi identity. It is convenient also to express the dual multiplet $P_L \Lambda_{D\,\alpha}$ as 
\be
P_L \Lambda_{D\,\alpha} = \Sigma (\mathcal{D}_\alpha U),
\ee
where $\mathcal{D}_\alpha$ is an operator, analogous to the superspace derivative~\cite{Kugo:1983mv}, that maps 
\be
\mathcal{D}_\alpha :(w,c)\to\left(w+\frac12,c-\frac32\right)
\ee
and $U$ is a vector multiplet with vanishing weights. Notice that, by using the properties of the $\Sigma$ and $\mathcal{D}_\alpha$ operators, up to boundary terms we have
\be
\begin{aligned}
\left[i \bar \Lambda_D P_L \Lambda\right]_F &= \left[\rmi \Sigma(\mathcal{D}^\alpha U P_L\Lambda_\alpha) \right]_F\\
& =\left[\rmi \mathcal{D}^\alpha U P_L\Lambda_\alpha-\rmi  \bar{\mathcal{D}}^\alpha U P_R\Lambda_\alpha\right]_D = -\left[\rmi U \mathcal{D}^\alpha P_L \Lambda_\alpha-\rmi U\bar{\mathcal{D}}^\alpha P_R \Lambda_\alpha\right]_D.
\end{aligned}
\ee

As a consistency check, we verify first that by integrating out $U$ we get back the original action \eqref{SV}. The variation of $U$ gives 
\be
\delta U: \qquad \mathcal{D}^\alpha P_L\Lambda_\alpha = \bar{\mathcal{D}}^\alpha P_R \Lambda_\alpha,
\ee
which is the supersymmetric form of the Bianchi identities implying that $P_L\Lambda_\alpha$ is the field strength of a vector multiplet V: $P_L\Lambda_\alpha=\Sigma( \mathcal{D}_\alpha V)$. Inserting this result in \eqref{tildeSV}, the action reduces therefore correctly to \eqref{SV}. On the other hand, by integrating out the Lagrange multiplier $P_L \Phi_\alpha$ and the unconstrained chiral multiplet $P_L \Lambda_\alpha$ we obtain
\begin{align}
\delta P_L \Phi_\alpha: &\qquad P_L \Lambda_\alpha X =0,\\
\label{eq:Wdual}
\delta P_L \Lambda_\alpha: &\qquad  \hat f_{\Db} P_L \Lambda_\alpha = \rmi P_L \Lambda_{D\,\alpha} - P_L \Phi_\alpha X.
\end{align}
By multiplying the second equation by $X$ and using the assumption that $X$ is nilpotent, we obtain the additional on-shell information
\be
\label{constrWDX}
P_L \Lambda_{D\,\alpha}X =0,
\ee
which means that the fermion in the dual multiplet $P_L \Lambda_{D\,\alpha}$ is removed. Inserting \eqref{eq:Wdual} back into the action \eqref{tildeSV} and using also \eqref{constrWDX} we obtain eventually the on-shell expression
\be
\tilde S_V = -\frac14 \left[\hat f_{\Db}^{-1}\bar \Lambda_D P_L \Lambda_D\right]_F.
\ee
Thus we have shown that the vector multiplet sector of the anti-D3-brane action enjoys self-duality. We notice finally that, due to the nilpotent constraint on $X$, $\hat f_{\Db}$ satisfies the property 
\be
(\hat f_{\Db} (\bar f))^{-1} =\hat f_{\Db} (\bar f^{-1}),
\ee
indeed
\be 
(\hat f_{\Db} (\bar f))^{-1} = \Sigma \left(\frac{\bar X^0 e^{-\frac K6}\bar X}{\Sigma\left(\bar X^0 e^{-\frac K6}\bar X \bar f\right)}\right) =\Sigma \left(\frac{\bar X^0 e^{-\frac K6}\bar X}{\Sigma\left(\bar X^0 e^{-\frac K6}\bar X\right)\bar f}\right) =  \hat f_{\Db} (\bar f^{-1}).
\ee
As a consequence, for the particular choice $f(\tau)=-\rmi \tau$, the transformation $\hat f_{\Db} \to  (\hat f_{\Db})^{-1}$ corresponds to $\tau \to -1/\tau$. This, together with $\tau \to \tau +1$ which is a trivial symmetry of the action, generates the full SL$(2,\mathbb{R})$ group.

\section{Conclusion}\label{sec:conclusion}

In this paper we studied the low energy effective action for the KKLT scenario~\cite{Kachru:2003aw}. In particular, we kept track of all world volume fields on the anti-D3-brane, i.e. the vector field, the scalars as well as the fermions. We showed, following~\cite{McGuirk:2012sb}, how the corresponding world volume fields couple to closed string moduli, i.e. the axio-dilaton, the complex structure moduli and the single K\"ahler modulus. We then rewrote this 4d effective action in a manifestly supersymmetric way, making use of constrained multiplets \cite{Komargodski:2009rz} that have played an important role in recent advances in our understanding of supergravity (see for example~\cite{Dudas:2015eha, Dall'Agata:2015lek, Schillo:2015ssx, Yamada:2016tca, Cribiori:2017ngp, Cribiori:2019cgz}).

Our manifestly supersymmetric 4d $\N=1$ action shows that the anti-D3-brane in KKLT breaks supersymmetry spontaneously. It also provides a useful reformulation of the anti-D3-brane action that should facilitate future more phenomenological studies of anti-branes in string compactifications. It furthermore goes beyond what was in the literature in an important way. Indeed, initial studies of the uplift term in KKLT in supergravity have focused on the Goldstino that can be packaged into a nilpotent chiral multiplet~\cite{Kallosh:2014wsa, Bergshoeff:2015jxa}, but the bosonic world volume fields on the anti-D3-brane have been neglected in this analysis because they can be projected out by placing the anti-D3-brane on top of an O3-plane~\cite{Kallosh:2015nia, Aparicio:2015psl, Garcia-Etxebarria:2015lif}. This work cumulated in~\cite{GarciadelMoral:2017vnz}, where the full action for all world volume fermions, including their couplings to the closed string moduli has been derived.

When adding the bosonic world volume fields, we followed the proposal of~\cite{Vercnocke:2016fbt, Kallosh:2016aep} of how one can package the bosons into constrained $\N=1$ multiplets. While this choice might not be unique it allowed us to rewrite the action in a manifestly supersymmetric way. One particular challenge one faces is related to the U$(1)$ gauge field on the anti-D3-brane. Recall that the gauge kinetic function $f$ for a D3-brane in the KKLT background is given by the axion dilaton $f(\tau)=-\rmi \, \tau = -\rmi (C_0 + \rmi e^{-\phi})$. This is and has to be a holomorphic function. However, the anti-D3-brane has the opposite sign in the Wess-Zumino part of its action and therefore the U$(1)$ world volume gauge field naturally couples to $\rmi\, \bar \tau =-\rmi( - C_0 + \rmi e^{-\phi})$. The latter is anti-holomorphic and such a gauge kinetic function is forbidden by supersymmetry. As explained in section~\ref{sec:sugraaction}, we resolve this puzzle by using a construction inspired by the recently discovered new D-term in supergravity~\cite{Cribiori:2017laj}.

While our paper completes the study of a single anti-D3-brane in the KKLT setup, there are a variety of future research directions that should be pursued: 
\begin{itemize}
\item It should be straightforward to generalize our results to a stack of $N_{\rm \overline{D3}}$ anti-D3-branes by simply promoting the world volume fields to fields that transform in the adjoint of SU$(N_{\rm \overline{D3}})$. This should amount to inserting traces into our formulas and adding commutator terms that appear in general~\cite{McGuirk:2012sb} but vanish in the abelian case. Note however, that such a stack of anti-D3-branes in the KKLT background would want to polarize into an NS5-brane~\cite{Kachru:2002gs}. Two recent papers~\cite{Aalsma:2017ulu, Aalsma:2018pll} studied the effective NS5-brane theory that arises from a stack of anti-D3-branes in the Klebanov-Strassler throat geometry~\cite{Klebanov:2000hb}. The authors showed that in the metastable minimum supersymmetry is non-linearly realized. This non-linearly realized and spontaneously broken supersymmetry can be linearly realized, if one includes a tower of massive Kaluza-Klein states~\cite{Aalsma:2018pll}. It would be interesting to study the polarization process in more detail and include all world volume fields on the anti-D3-branes.
\item Since a reformulation in terms of constrained multiplets is not necessarily unique, one should explore other supersymmetric formulations of the action. For example, in \cite{Bandos:2016xyu, GarciadelMoral:2017vnz} the authors discuss also a constrained vector superfield that contains the Goldstino as the only degree of freedom. This vector superfield can replace the nilpotent chiral superfield and generates the KKLT uplift term via a D-term. It is conceivable that for the other constrained multiplets there are also alternative formulations that should be explored.
\item Our results should be extendible  to the Large Volume Scenario~\cite{Balasubramanian:2005zx, Conlon:2005ki}. There the AdS vacuum breaks supersymmetry already before adding the uplifting anti-D3-brane. This means that the Goldstino is always a combination of closed string fermions and the fermions on the anti-D3-brane. Nevertheless, one should be able to describe the anti-D3-brane action in the same way that we did here. In particular, the nilpotent chiral superfield is still consisting of only a world volume fermion.
\item Recently it was shown that the anti-D3-brane in the KKLT scenario is one particular case of flux compactifications with anti-Dp-branes that can all be described by a 4d effective $\N=1$ supergravity action that includes a nilpotent chiral multiplet \cite{Kallosh:2018nrk}. It would be interesting to work out the proper description of the world volume fields for anti-Dp-branes with $p>3$. While this is not easy, one should be able to adapt existing results for the light degrees of freedom on supersymmetric branes (see for example~\cite{Jockers:2004yj, Grimm:2011dx, Kerstan:2011dy, Carta:2016ynn, Escobar:2018tiu}) since the DBI action for branes and anti-branes is the same.
\item It would be very interesting to derive the brane action in a non-trivial background beyond quadratic order in the fermions. In particular, the quartic terms play an important role in the study of the 10d lift of the 4d KKLT solution~\cite{Moritz:2017xto, Moritz:2018sui, Kallosh:2018wme, Moritz:2018ani, Kallosh:2018psh, Gautason:2018gln, Hamada:2018qef, Kallosh:2019axr, Kallosh:2019oxv, Hamada:2019ack, Carta:2019rhx, Gautason:2019jwq}. While technically challenging to obtain, these higher order fermionic terms would have applications well beyond the study of anti-D-branes in flux compactifications.
\end{itemize}

We hope to study some of these issues in the future.

\acknowledgments
We are grateful to Fotis Farakos, Renata Kallosh, Liam McAllister, Susha Parameswaran, Augusto Sagnotti, Harald Skarke, Flavio Tonioni, Magnus Tournoy and Antoine van Proeyen for enlightening discussions and collaboration on related projects and to Liam McAllister for comments on an earlier version of this draft. The work of NC, CR and TW is supported by an FWF grant with the number P 30265. The work of YY was supported by the SITP, the US National Science Foundation grant PHY-1720397, and Grant-in-Aid for JSPS Fellows (19J00494).

\appendix

\section{Superspace}

In this appendix we translate the relevant formulae into the language of superspace, following the conventions of~\cite{Wess:1992cp}. In this case fermions are described by two-components Weyl spinors. We recall that a four-component spinor can be written in terms of a two-components one as $\Omega = \left(\begin{array}{c}\psi\\\bar\psi\end{array}\right)$.

\subsection{Constrained superfields}\label{appA:superspace}
We describe here constrained superfields in flat superspace.
\begin{itemize}
\item The nilpotent chiral Goldstino superfield is given by
\be
X = \frac{G^2}{2F} + \sqrt 2 \theta G + \theta^2 F \qquad \Leftrightarrow \qquad X^2=0,
\ee
where $G_\alpha$ is the spin-1/2 Goldstino. Upon substituting $\theta\to \Theta$, this expression is valid also in supergravity.

\item A chiral superfield $Y$, such that $XY=0$ is instead
\be
Y = \frac{G\chi}{F}-\frac{G^2}{2F^2}F^Y + \sqrt 2 \theta \chi + \theta^2 F^Y.
\ee
Upon substituting $\theta\to \Theta$, this expression is valid also in supergravity.

\item A chiral superfield $H$ 
\be
H = h + \sqrt{2} \theta \chi^H +\theta^2 F^H
\ee
containing only the complex scalar in the lowest component is subjected to the constraint
\be
X \bar D_{\dot \alpha} \bar H =0,
\ee
where $X^2=0$ is assumed. Its fermion and its auxiliary field are given by
\begin{align}
\chi^H &= i \sigma^\mu \left(\frac{\bar{G}}{\bar{F}} \right)\partial_\mu h,\\
F^H &= -\partial_\mu \left(\frac{\bar{G}}{\bar{F}}\right)\bar{\sigma}^\nu \sigma^\mu \frac{\bar{G}}{\bar{F}} \partial_\nu h + \frac{\bar{G}^2}{2\bar{F}^2}  \Box h.
\end{align}
The generalization to supergravity can be found in~\cite{Dall'Agata:2015lek}.

\item Given a real abelian vector superfield $V$, the field strength chiral superfield is 
\be 
W_\alpha =- \frac{1}{4} \bar{D}^2 D_\alpha V = -\rmi \Lambda_\alpha + L^\beta_{\alpha} \theta_\beta + \sigma^\mu_{\alpha \dot{\alpha}}\partial_\mu \bar{\Lambda}^{\dot{\alpha}} \theta^2\,,
\ee
where
\be 
L^\beta_\alpha = \delta^\beta_\alpha D - \frac{\rmi}{2} \left(\sigma^\mu \sigma^\nu\right)^\beta_\alpha F_{\mu\nu}.
\ee
The gaugino $\Lambda_\alpha$ can be removed via the constraint
\be 
X W_\alpha = 0,
\ee
which leads to
\be 
\begin{aligned}
\Lambda_\alpha &= \frac{\rmi}{\sqrt 2 F} L_\alpha^\beta G_\beta 
- \frac{G^2}{2F^2}\partial_\mu \left(\frac{\bar G_{\bar \dot \alpha}\bar L^{\dot\alpha}_{\dot\beta}}{\sqrt 2 \bar F}\right)\bar \sigma^{\mu\, \dot{\beta}{\gamma}}\epsilon_{\gamma\alpha}\\
&-\rmi\frac{G^2}{2F^2}\sigma^{\mu}_{\alpha\dot\beta}\bar\sigma^{\nu\,\dot\beta\gamma}\partial_\mu\left(\frac{\bar G^2}{2\bar F^2}\partial_\nu \left(\frac{L_\gamma^\delta G_\delta}{\sqrt 2F}\right)\right)\\
&-\frac12 \frac{G^2}{F^2}\frac{\bar G^2}{\bar F^2}\left(\partial \left(\frac{G}{\sqrt 2 F}\right)\right)^2 \partial_\mu \left(\frac{\bar G_{\dot\alpha}\bar L^{\dot\alpha}_{\dot\beta}}{\sqrt 2 \bar F}\right)\bar \sigma^{\mu \, \dot \beta \gamma}\epsilon_{\gamma\alpha}\,.
\end{aligned}
\ee
The supergravity expression for the removed gaugino can be found in~\cite{Dall'Agata:2015lek}.

\end{itemize}

\subsection{The anti-D3-brane supergravity Lagrangian}
The supergravity Lagrangian for the anti-D3-brane can be expressed in superspace as
\be
\mathcal{L} = -3 \int d^4 \theta E e^{-\frac K3} + \frac14 \left(\int d^2 \Theta 2 \mathcal{E} \hat f_{\Db} \mathcal{W}^\alpha \mathcal{W}_\alpha +c.c.\right) + \left(\int d^2 \Theta 2 \mathcal{E} {W} +c.c.\right), 
\ee
where $K$ is the \K potential, $W$ the superpotential and $\mathcal{W}_\alpha = -\frac14 (\bar{\mathcal{D}}^2-8 \mathcal{R})\mathcal{D}_\alpha V$ is the field strength superfield. The anti-D3-brane gauge kinetic function is
\be
\label{fD3superspace}
\hat f_{\Db} = (\bar{\mathcal{D}}^2-8 \mathcal{R})\left(\frac{\bar X \bar f}{\bar{\mathcal{D}}^2\bar X}\right),
\ee
where $f(\tau) = -\rmi \tau$. The expressions \eqref{eq:Kfinal1}, \eqref{eq:Wfinal1} and \eqref{eq:Kfinal2}, \eqref{eq:Wfinal2} we gave for $K$ and $W$ can be used directly without any modification. When using  the second alternative, the superpotential mass for the fermions in $Y^i$ is generated by the chiral superfield
\be
\label{MD3superspace}
\hat M_{ij}  = (\bar{\mathcal{D}}^2-8 \mathcal{R})\left(\frac{e^{-4\A-2u}\bar X m_{ij}}{\bar{\mathcal{D}}^2\bar X\, (-\rmi(\tau-\bar\tau))^\frac12 f(U^A,\bar U^A)^{-\frac16}}\right),
\ee
where $m_{ij}$ is given in \eqref{mij}.

We notice finally that the expressions \eqref{fD3superspace} and \eqref{MD3superspace} can be written in an alternative form, that may be preferred in some applications. Given a nilpotent Goldstino superfield $X$, we can define a superfield \cite{Cribiori:2016hdz,Bandos:2016xyu}
\be
\Gamma_\alpha = -2\sqrt 2 \frac{\mathcal{D}_\alpha X}{\mathcal{D}^2 X}.
\ee
It is possible to check then that such a superfield satisfies
\begin{equation}
\begin{aligned}
\mathcal{D}_\beta \Gamma_\alpha &= \sqrt 2\epsilon_{\alpha \beta}\left(1- \bar{\mathcal{R}} \,\Gamma^2\right),\\
\bar{\mathcal{D}}^{\dot \beta} \Gamma^\alpha &= \sqrt 2 \rmi (\bar \sigma^b \Gamma)^{\dot \beta}\mathcal{D}_b \Gamma^\alpha+\frac{1}{2 \sqrt 2} \Gamma^2 B^{\dot \beta \alpha},
\end{aligned}
\end{equation}
where $\mathcal{D}_a \Gamma_\alpha = e^m_b \mathcal{D}_m \Gamma_\alpha-\frac12 \psi_a^\beta \mathcal{D}_\beta \Gamma_\alpha -\frac12 \bar \psi_{b \dot \beta}\bar{\mathcal{D}}^{\dot \beta}\Gamma_\alpha$ is the supercovariant derivative in superspace and the definition of the superfield $B_{\alpha \dot \beta}$ can be found in \cite{Wess:1992cp}. These are the conditions given in \cite{Samuel:1982uh} in order for $\Gamma_\alpha$ to describe a Goldstino. Indeed the lowest component $\Gamma_\alpha|=\gamma_\alpha$, is the (chiral) Volkov--Akulov Goldstino, while all of the other fields inside $\Gamma_\alpha$ are removed. With this new ingredient at our disposal, the gauge kinetic function and the superpotential mass can be expressed in a more compact form as
\be
\begin{aligned}
\hat f_{\Db} &=- \frac{1}{8} (\bar{\mathcal{D}}^2-8 \mathcal{R})(\bar \Gamma^2 \bar f),\\
\hat M_{ij} &= -\frac{1}{8} (\bar{\mathcal{D}}^2-8 \mathcal{R})\left(\bar \Gamma^2 \frac{e^{-4\A-2u}}{(-\rmi (\tau-\bar \tau))^\frac12 f(U^A,\bar U^A)^{-\frac16} }m_{ij}\right). 
\end{aligned}
\ee
Their lowest components in the unitary gauge $\gamma_\alpha=0$ are
\be
\hat f_{\Db}| = \bar f, \qquad \hat M_{ij}| = \frac{e^{-4\A-2u}}{(-\rmi (\tau-\bar \tau))^\frac12 f(U^A,\bar U^A)^{-\frac16} }m_{ij}.
\ee
We stress that such a construction is general, namely given an arbitrary superfield $\Phi$, which can also be composite, the chiral superfield
\be
\hat \Phi = -\frac18  (\bar{\mathcal{D}}^2-8 \mathcal{R})(\bar \Gamma^2 \Phi)
\ee
has lowest component
\be
\hat \Phi| = \Phi+\text{fermions}
\ee
and the fermionic terms vanish in the unitary gauge. It satisfies the constraint
\be
\Gamma^2 \bar \Gamma^2 \hat \Phi = \Gamma^2 \bar \Gamma^2 \Phi.   
\ee
For completeness, we give the components of $\hat f_{\Db}$ in global supersymmetry. The chiral superfield $\hat f_{\Db}$ can be expanded in superspace as
\be
\hat f_{\Db} = \hat f_{\Db}| + \theta^\alpha D_\alpha \hat f_{\Db}| -\frac14 \theta^2 D^2 \hat f_{\Db}|,
\ee 
where 
\begin{align}
\hat f_{\Db}| &=\bar f + \frac{2\bar D_{\dot\alpha}\bar X \bar D^{\dot \alpha}\bar f + \bar X \bar D^2 \bar f}{\bar D^2 \bar X},\\
D_\alpha \hat f_{\Db}| &=-4\rmi \partial_{\alpha\dot\alpha}\bar X\frac{\bar D^{\dot\alpha}\bar f}{\bar D^2 \bar X}-4\rmi \partial_{\alpha\dot\alpha}\bar f\frac{\bar D^{\dot\alpha}\bar X}{\bar D^2 \bar X}-4\rmi \frac{\bar X}{\bar D^2 \bar X}\partial_{\alpha\dot\alpha}\bar D^{\dot \beta}\bar f\\
&\nn+4\rmi \partial_{\alpha\dot\alpha}\bar D^{\dot \alpha}\bar X\left(\frac{2 \bar D_{\dot\beta}\bar X \bar D^{\dot\beta}\bar f + \bar X \bar D^2 \bar f}{(\bar D^2 \bar X)^2}\right),\\
D^2 \hat f_{\Db}| &= -16 \frac{\partial_{\alpha\dot\alpha}\bar X \partial^{\dot\alpha\alpha}\bar f}{\bar D^2 \bar X}-32 \frac{\partial_{\alpha\dot\alpha}\bar X \bar D^{\dot\alpha}\bar f \partial^{\dot\beta \alpha}\bar D_{\dot\beta}\bar X}{(\bar D^2\bar X)^2}-32 \frac{\partial_{\alpha\dot\alpha}\bar f \bar D^{\dot\alpha}\bar X \partial^{\dot\beta \alpha}\bar D_{\dot\beta}\bar X}{(\bar D^2\bar X)^2}\nn\\
&+16 \frac{\bar X \square \bar f}{\bar D^2 \bar X}-32 \frac{\bar X \partial^{\dot\beta \alpha}\bar D_{\dot\beta}\bar X\partial_{\alpha\dot\gamma}\bar D^{\dot\gamma}\bar f}{(\bar D^2 \bar X)^2}-16\frac{\square \bar X\left(2 \bar D_{\dot\alpha}\bar X \bar D^{\dot\alpha}\bar f + \bar X \bar D^2 \bar f\right)}{(\bar D^2\bar X)^2}\\
&\nn-64 \frac{\partial_{\alpha \dot\beta}\bar D^{\dot\beta}\bar X \bar D_{\dot\gamma}\bar X \bar D^{\dot \gamma}\bar f \partial^{\dot\alpha \alpha}\bar D_{\dot \alpha}\bar X}{(\bar D^2 \bar X)^3}-32 \frac{\bar X\partial_{\alpha\dot\beta}\bar D^{\dot\beta}\bar X\bar D^2 \bar f \partial^{\dot\alpha\alpha}\bar D_{\dot\alpha }\bar X}{(\bar D^2 \bar X)^3}.
\end{align}
The projection to $\theta=0$ on the right hand sides of the expressions above is understood and $X^2=0$ is always assumed.

\section{New D-terms in supergravity and the theta term}\label{appB:newDterm}

In this appendix we review some ingredients concerning the construction of the new Fayet--Iliopoulos D-terms in supergravity originally proposed in~\cite{Cribiori:2017laj}. This discussion is needed to understand the origin of the coupling \eqref{eq:vector}, which we introduced in order to flip the sign of the theta term in the supergravity action.

Let us start by considering the chiral field strength multiplet $P_L\Lambda_{\alpha}$ given in~\eqref{Lambdamult} and with weight $\frac32$. In particular, $P_L\Lambda_{\alpha}$ is the gaugino, $F_{ab}$ is the covariant field strength of the U$(1)$ vector and D is the real auxiliary field. It is known that the standard embedding of a Fayet--Iliopoulos term 
\be
\label{LFI}
S_{FI} = -\xi\int d^4 x \sqrt{-g_4} \,{\rm D}
\ee
in supergravity requires the gauging of the U$(1)$ R-symmetry, by means of the vector field. In~\cite{Cribiori:2017laj} a new embedding has been proposed that avoids such a restriction and that is of the type
\be
\label{newDterm}
S_\text{FI new} =-\xi \bigg[\frac{\omega^2 \bar \omega^2}{\Sigma(\bar \omega^2)\bar \Sigma(\omega^2)}X^0 \bar X^0 \, {\rm D}\bigg]_D = -\xi \int d^4x \sqrt{-g_4}\,X^0 \bar X^0 {\rm D} + \dots,
\ee
where dots stand for fermionic terms. In this expression, $X^0$ is the compensator chiral multiplet, ${\rm D}$ is a real multiplet which has the auxiliary field in the lowest component (see formula (3.3) of~\cite{Cribiori:2017laj}) and we defined the multiplets
\be
\omega^2 = \frac{\bar \Lambda P_L \Lambda}{(X^0\bar X^0)^2}, \qquad \bar\omega^2 = \frac{\bar \Lambda P_R \Lambda}{(X^0\bar X^0)^2}.
\ee
$\Sigma$ and $\bar \Sigma$ are the superconformal generalizations of the chiral projectors $\bar D^2$ and $D^2$ of superspace. They act on multiplets with weights $(\text{Weyl}, \text{chiral})=(w,\pm (w-2))$ and produce respectively chiral and anti-chiral multiplets, namely 
\be
\Sigma:\, (w,w-2) \to (w+1,w+1),\qquad \bar \Sigma:\,(w,-w+2) \to (w+1,-w-1).
\ee
More details about these operators can be found for example in~\cite{Ferrara:2016een}, where they are denoted with $T$ and $\bar T$. We need also the components of the chiral multiplet
\begin{equation}
\bar \Lambda P_L\Lambda = \{\bar \Lambda P_L \Lambda, \sqrt{2}P_L\left(\rmi {\rm D} - \frac12 \slashed{\hat F}\right)\Lambda, 2 \bar \Lambda P_L \slashed{\mathcal{D}}\Lambda+\hat F^- \cdot \hat F^- - {\rm D}^2\},
\end{equation}
from which one can calculate
\begin{equation}
\bar \Sigma(\omega^2) = (X^0\bar X^0)^{-2}\left( \frac12 F_{\mu\nu}F^{\mu\nu}+\frac i4 \frac{\epsilon^{\mu\nu\rho\sigma}}{\sqrt{-g_4}}F_{\mu\nu}F_{\rho\sigma}-{\rm D}^2+\dots\right),
\end{equation}
where the dots stand for fermionic terms. We recall that we are following the notation and the conventions of \cite{freedman2012supergravity}.

The important property of the coupling~\eqref{newDterm} is that its pure bosonic sector contains just a term linear in the auxiliary field D, while all of the remaining fermionic terms are required by superconformal symmetry. After gauge fixing the superconformal symmetry to Poincar\'e supergravity, $X^0 =\kappa^{-1}e^{\frac K6}$, the new D-term reduces then to ($\kappa=1$)
\be
\label{SFINEW}
S_\text{FI new} = -\xi \int d^4 x \sqrt{-g_4}\,e^{\frac K3} {\rm D}+\dots
\ee
and therefore, when this is added to the standard supergravity action \eqref{LV} for the vector multiplet, the auxiliary field D acquires a non-vanishing vacuum expectation value and supersymmetry is spontaneously broken. Setting then the Goldstino to zero by a unitary gauge choice, all of the fermionic interactions in \eqref{SFINEW} vanish. Notice that the presence of this coupling spoils the \K invariance of the theory, since the \K potential is appearing explicitly in the action. The \K invariant version of~\eqref{newDterm} has been constructed in~\cite{Antoniadis:2018oeh} and it amounts to replacing $X^0 \bar X^0 \to X^0 \bar X^0 e^{-\frac K3}$ in order to cancel the undesired dependence on $K$ in Poincar\'e frame. In what follows we systematically perform this replacement.

The logic behind~\eqref{newDterm} is the following. First, notice that it is constructed out of the two real multiplets
\be
\label{R12}
R_1=\frac{\omega^2 \bar \omega^2}{\Sigma(\bar \omega^2)\bar \Sigma(\omega^2)}, \qquad R_2=X^0 \bar X^0 e^{-\frac K3}\, {\rm D},
\ee
with weights $(-2,0)$ and $(4,0)$ respectively, and that the component expansion starts precisely with the lowest component of $R_2$. This means that the role of $R_1$ is to provide higher order fermionic interactions which are needed in order to write down a consistent superconformal embedding of $R_2$. The procedure can be easily generalized in order to construct the superconformal completion of any arbitrary real multiplet $R_2$ with weights $(4,0)$. A similar logic has been followed in \cite{Farakos:2018sgq}, by using only chiral multiplets.

In section~\ref{subsec:vecsugra} we are interested in deforming a given supergravity action in a supersymmetric way with a theta term
\begin{equation}
\label{thetaterm}
S_\text{$\theta$-term}=-\frac14 \int d^4 x\,{\rm Im}(f)\epsilon^{\mu\nu\rho\sigma}F_{\mu\nu}F_{\rho\sigma}+\dots,
\end{equation}
where the dots are additional couplings which are needed from supersymmetry and which we would like to determine. The standard procedure to find them consists in varying \eqref{thetaterm} and adding the required terms in order to cancel the total variation and obtain a superconformal invariant coupling. This strategy is conceptually simple but tedious and it is not clear at which step it is going to end. However, when supersymmetry is spontaneously broken the problem can be solved at once by considering the multiplet $R_1$ as defined in \eqref{R12} and multiplying it by
\be
R_2={\rm Im}(f)\epsilon^{\mu\nu\rho\sigma}F_{\mu\nu}F_{\rho\sigma},
\ee
which has to be thought of as a real multiplet with weights (4,0). In particular, $f(\tau)=-\rmi \tau$ can be understood as the lowest component of a chiral multiplet with vanishing weight. This strategy is formally correct, but there is one additional subtlety one needs to be careful about in our case. In the anti-D3-brane action we are proposing, the auxiliary field of the vector multiplet is not acquiring a vacuum expectation value, since supersymmetry is spontaneously broken by the auxiliary field $F$ of $X$. As a consequence, we are not allowed to divide by the quantity $\bar \Sigma (\omega^2)$, as it vanishes in the vacuum. However, in the case in which the field strength multiplet is constrained as in \eqref{constrXW}, we can use the following identity proved in \cite{Cribiori:2017laj} 
\be
\frac{\omega^2 \bar \omega^2}{\Sigma(\bar \omega^2)\bar \Sigma (\omega^2)} = \frac{X^0 \bar X^0 X \bar X}{\Sigma (\bar X^0\bar X) \bar \Sigma(X^0 X)} = \frac{X^0 \bar X^0 e^{-\frac K3}\, X \bar X}{\Sigma (\bar X^0e^{-\frac K6}\bar X) \bar \Sigma(X^0e^{-\frac K6} X)} ,
\ee
to trade formally $\omega^2$ for the nilpotent $X$, inside $R_1$. In particular, the right hand side is well defined in our setup, since the denominator can never vanish. We are therefore in a position to propose the following superconformal invariant interaction
\be
\label{Ltheta}
\begin{aligned}
S_\text{$\theta$-term}=&-\frac14\bigg[\frac{X^0 \bar X^0 e^{-\frac K3}\, X \bar X}{\Sigma (\bar X^0e^{-\frac K6}\bar X) \bar \Sigma(X^0e^{-\frac K6} X)} \, {\rm Im}(f)\frac{\epsilon^{\mu\nu\rho\sigma}}{\sqrt{-g_4}}F_{\mu\nu}F_{\rho\sigma}\bigg]_D \\
=& -\frac14  {\rm Im}(f)\epsilon^{\mu\nu\rho\sigma}F_{\mu\nu}F_{\rho\sigma} + \dots,
\end{aligned}
\ee
which has the desired property of providing a consistent superconformal completion of  \eqref{thetaterm}, much in the same way that \eqref{newDterm} provides it for the Fayet--Iliopoulos term \eqref{LFI}. We can also express the multiplet $\epsilon^{\mu\nu\rho\sigma}F_{\mu\nu}F_{\rho\sigma}$ in terms of the more familiar field strength multiplet $P_L\Lambda$. One can indeed check that
\be
\left(X^0 \bar X^0 e^{-\frac K3}\right)^2 \frac{1}{2{\rm i}}\left[\bar \Sigma(\omega^2)- \Sigma(\bar \omega^2)\right] = \frac14 \frac{\epsilon^{\mu\nu\rho\sigma}}{\sqrt{-g_4}} F_{\mu\nu}F_{\rho\sigma} +\dots,
\ee
where the dots stand for fermionic terms containing at least one gaugino $\Lambda_\alpha$ not acted upon with derivatives. Due to the Grassmann nature of $\Lambda_\alpha$, these terms vanish if multiplied by $X \bar X$ and we have directly the constraint
\be
X \bar X\left(X^0 \bar X^0 e^{-\frac K3}\right)^2 \frac{1}{2{\rm i}}\left[\bar \Sigma(\omega^2)- \Sigma(\bar \omega^2)\right] =\frac14 X \bar X\frac{\epsilon^{\mu\nu\rho\sigma}}{\sqrt{-g_4}} F_{\mu\nu}F_{\rho\sigma},
\ee
where every quantity can be understood as a full supersymmetric multiplet. The proposed coupling~\eqref{Ltheta} becomes then
\be
\begin{aligned}
S_\text{$\theta$-term}=&-\frac14\bigg[\frac{X^0 \bar X^0 e^{-\frac K3}\, X \bar X}{\Sigma (\bar X^0e^{-\frac K6}\bar X) \bar \Sigma(X^0e^{-\frac K6} X)} \, {\rm Im}(f)\frac{\epsilon^{\mu\nu\rho\sigma}}{\sqrt{-g_4}}F_{\mu\nu}F_{\rho\sigma}\bigg]_D \\
=&-\frac{1}{2{\rm i}}\bigg[\frac{ X \bar X}{\Sigma (\bar X^0e^{-\frac K6}\bar X) \bar \Sigma(X^0e^{-\frac K6} X)}\left(X^0 \bar X^0 e^{-\frac K3}\right)^3\, {\rm Im}(f)\left(\bar \Sigma(\omega^2)- \Sigma(\bar \omega^2)\right)\bigg]_D \\
=&\bigg[\frac{X \bar X}{\Sigma (\bar X^0e^{-\frac K6}\bar X) \bar \Sigma(X^0e^{-\frac K6} X)}\left(X^0 \bar X^0 e^{-\frac K3}\right)^3\, {\rm Im}(f)\,{\rm Im}\left(\Sigma(\bar \omega^2)\right)\bigg]_D\\
=& -\frac14 \int d^4 x \, {\rm Im}(f)\epsilon^{\mu\nu\rho\sigma}F_{\mu\nu}F_{\rho\sigma}+\dots,
\end{aligned}
\ee
which is the expression appearing in~\eqref{eq:vector}.

\bibliographystyle{JHEP}

\providecommand{\href}[2]{#2}\begingroup\raggedright\begin{thebibliography}{}

\end{thebibliography}\endgroup


\begin{thebibliography}{100}

\bibitem{Kachru:2003aw}
S.~Kachru, R.~Kallosh, A.~D. Linde and S.~P. Trivedi, \emph{{De Sitter vacua in
  string theory}},
  \href{https://doi.org/10.1103/PhysRevD.68.046005}{\emph{Phys. Rev.}
  {\bfseries D68} (2003) 046005}
  [\href{https://arxiv.org/abs/hep-th/0301240}{{\ttfamily hep-th/0301240}}].

\bibitem{Brennan:2017rbf}
T.~D. Brennan, F.~Carta and C.~Vafa, \emph{{The String Landscape, the
  Swampland, and the Missing Corner}},
  \href{https://doi.org/10.22323/1.305.0015}{\emph{PoS} {\bfseries TASI2017}
  (2017) 015} [\href{https://arxiv.org/abs/1711.00864}{{\ttfamily
  1711.00864}}].

\bibitem{Danielsson:2018ztv}
U.~H. Danielsson and T.~Van~Riet, \emph{{What if string theory has no de Sitter
  vacua?}}, \href{https://doi.org/10.1142/S0218271818300070}{\emph{Int. J. Mod.
  Phys.} {\bfseries D27} (2018) 1830007}
  [\href{https://arxiv.org/abs/1804.01120}{{\ttfamily 1804.01120}}].

\bibitem{Palti:2019pca}
E.~Palti, \emph{{The Swampland: Introduction and Review}},  2019,
  \href{https://arxiv.org/abs/1903.06239}{{\ttfamily 1903.06239}}.

\bibitem{Kachru:2002gs}
S.~Kachru, J.~Pearson and H.~L. Verlinde, \emph{{Brane / flux annihilation and
  the string dual of a nonsupersymmetric field theory}},
  \href{https://doi.org/10.1088/1126-6708/2002/06/021}{\emph{JHEP} {\bfseries
  06} (2002) 021} [\href{https://arxiv.org/abs/hep-th/0112197}{{\ttfamily
  hep-th/0112197}}].

\bibitem{Ferrara:2014kva}
S.~Ferrara, R.~Kallosh and A.~Linde, \emph{{Cosmology with Nilpotent
  Superfields}}, \href{https://doi.org/10.1007/JHEP10(2014)143}{\emph{JHEP}
  {\bfseries 10} (2014) 143} [\href{https://arxiv.org/abs/1408.4096}{{\ttfamily
  1408.4096}}].

\bibitem{Kallosh:2014wsa}
R.~Kallosh and T.~Wrase, \emph{{Emergence of Spontaneously Broken Supersymmetry
  on an Anti-D3-Brane in KKLT dS Vacua}},
  \href{https://doi.org/10.1007/JHEP12(2014)117}{\emph{JHEP} {\bfseries 12}
  (2014) 117} [\href{https://arxiv.org/abs/1411.1121}{{\ttfamily 1411.1121}}].

\bibitem{Bergshoeff:2015jxa}
E.~A. Bergshoeff, K.~Dasgupta, R.~Kallosh, A.~Van~Proeyen and T.~Wrase,
  \emph{{$ \overline{\mathrm{D}3} $ and dS}},
  \href{https://doi.org/10.1007/JHEP05(2015)058}{\emph{JHEP} {\bfseries 05}
  (2015) 058} [\href{https://arxiv.org/abs/1502.07627}{{\ttfamily
  1502.07627}}].

\bibitem{Kallosh:2015nia}
R.~Kallosh, F.~Quevedo and A.~M. Uranga, \emph{{String Theory Realizations of
  the Nilpotent Goldstino}},
  \href{https://doi.org/10.1007/JHEP12(2015)039}{\emph{JHEP} {\bfseries 12}
  (2015) 039} [\href{https://arxiv.org/abs/1507.07556}{{\ttfamily
  1507.07556}}].

\bibitem{Garcia-Etxebarria:2015lif}
I.~Garc\'ia-Etxebarria, F.~Quevedo and R.~Valandro, \emph{{Global String
  Embeddings for the Nilpotent Goldstino}},
  \href{https://doi.org/10.1007/JHEP02(2016)148}{\emph{JHEP} {\bfseries 02}
  (2016) 148} [\href{https://arxiv.org/abs/1512.06926}{{\ttfamily
  1512.06926}}].

\bibitem{Sugimoto:1999tx}
S.~Sugimoto, \emph{{Anomaly cancellations in type I D-9 - anti-D-9 system and
  the USp(32) string theory}},
  \href{https://doi.org/10.1143/PTP.102.685}{\emph{Prog. Theor. Phys.}
  {\bfseries 102} (1999) 685}
  [\href{https://arxiv.org/abs/hep-th/9905159}{{\ttfamily hep-th/9905159}}].

\bibitem{Antoniadis:1999xk}
I.~Antoniadis, E.~Dudas and A.~Sagnotti, \emph{{Brane supersymmetry breaking}},
  \href{https://doi.org/10.1016/S0370-2693(99)01023-0}{\emph{Phys. Lett.}
  {\bfseries B464} (1999) 38}
  [\href{https://arxiv.org/abs/hep-th/9908023}{{\ttfamily hep-th/9908023}}].

\bibitem{Angelantonj:1999jh}
C.~Angelantonj, \emph{{Comments on open string orbifolds with a nonvanishing
  B(ab)}}, \href{https://doi.org/10.1016/S0550-3213(99)00662-8}{\emph{Nucl.
  Phys.} {\bfseries B566} (2000) 126}
  [\href{https://arxiv.org/abs/hep-th/9908064}{{\ttfamily hep-th/9908064}}].

\bibitem{Aldazabal:1999jr}
G.~Aldazabal and A.~M. Uranga, \emph{{Tachyon free nonsupersymmetric type IIB
  orientifolds via Brane - anti-brane systems}},
  \href{https://doi.org/10.1088/1126-6708/1999/10/024}{\emph{JHEP} {\bfseries
  10} (1999) 024} [\href{https://arxiv.org/abs/hep-th/9908072}{{\ttfamily
  hep-th/9908072}}].

\bibitem{Angelantonj:1999ms}
C.~Angelantonj, I.~Antoniadis, G.~D'Appollonio, E.~Dudas and A.~Sagnotti,
  \emph{{Type I vacua with brane supersymmetry breaking}},
  \href{https://doi.org/10.1016/S0550-3213(00)00052-3}{\emph{Nucl. Phys.}
  {\bfseries B572} (2000) 36}
  [\href{https://arxiv.org/abs/hep-th/9911081}{{\ttfamily hep-th/9911081}}].

\bibitem{Dudas:2000nv}
E.~Dudas and J.~Mourad, \emph{{Consistent gravitino couplings in
  nonsupersymmetric strings}},
  \href{https://doi.org/10.1016/S0370-2693(01)00777-8}{\emph{Phys. Lett.}
  {\bfseries B514} (2001) 173}
  [\href{https://arxiv.org/abs/hep-th/0012071}{{\ttfamily hep-th/0012071}}].

\bibitem{Pradisi:2001yv}
G.~Pradisi and F.~Riccioni, \emph{{Geometric couplings and brane supersymmetry
  breaking}}, \href{https://doi.org/10.1016/S0550-3213(01)00441-2}{\emph{Nucl.
  Phys.} {\bfseries B615} (2001) 33}
  [\href{https://arxiv.org/abs/hep-th/0107090}{{\ttfamily hep-th/0107090}}].

\bibitem{Bertolini:2015hua}
M.~Bertolini, D.~Musso, I.~Papadimitriou and H.~Raj, \emph{{A goldstino at the
  bottom of the cascade}},
  \href{https://doi.org/10.1007/JHEP11(2015)184}{\emph{JHEP} {\bfseries 11}
  (2015) 184} [\href{https://arxiv.org/abs/1509.03594}{{\ttfamily
  1509.03594}}].

\bibitem{Bandos:2015xnf}
I.~Bandos, L.~Martucci, D.~Sorokin and M.~Tonin, \emph{{Brane induced
  supersymmetry breaking and de Sitter supergravity}},
  \href{https://doi.org/10.1007/JHEP02(2016)080}{\emph{JHEP} {\bfseries 02}
  (2016) 080} [\href{https://arxiv.org/abs/1511.03024}{{\ttfamily
  1511.03024}}].

\bibitem{Dasgupta:2016prs}
K.~Dasgupta, M.~Emelin and E.~McDonough, \emph{{Fermions on the antibrane:
  Higher order interactions and spontaneously broken supersymmetry}},
  \href{https://doi.org/10.1103/PhysRevD.95.026003}{\emph{Phys. Rev.}
  {\bfseries D95} (2017) 026003}
  [\href{https://arxiv.org/abs/1601.03409}{{\ttfamily 1601.03409}}].

\bibitem{Vercnocke:2016fbt}
B.~Vercnocke and T.~Wrase, \emph{{Constrained superfields from an anti-D3-brane
  in KKLT}}, \href{https://doi.org/10.1007/JHEP08(2016)132}{\emph{JHEP}
  {\bfseries 08} (2016) 132}
  [\href{https://arxiv.org/abs/1605.03961}{{\ttfamily 1605.03961}}].

\bibitem{Kallosh:2016aep}
R.~Kallosh, B.~Vercnocke and T.~Wrase, \emph{{String Theory Origin of
  Constrained Multiplets}},
  \href{https://doi.org/10.1007/JHEP09(2016)063}{\emph{JHEP} {\bfseries 09}
  (2016) 063} [\href{https://arxiv.org/abs/1606.09245}{{\ttfamily
  1606.09245}}].

\bibitem{Bandos:2016xyu}
I.~Bandos, M.~Heller, S.~M. Kuzenko, L.~Martucci and D.~Sorokin, \emph{{The
  Goldstino brane, the constrained superfields and matter in $ \mathcal{N}=1 $
  supergravity}}, \href{https://doi.org/10.1007/JHEP11(2016)109}{\emph{JHEP}
  {\bfseries 11} (2016) 109}
  [\href{https://arxiv.org/abs/1608.05908}{{\ttfamily 1608.05908}}].

\bibitem{Aoki:2016tod}
S.~Aoki and Y.~Yamada, \emph{{More on DBI action in 4D $ \mathcal{N} $ = 1
  supergravity}}, \href{https://doi.org/10.1007/JHEP01(2017)121}{\emph{JHEP}
  {\bfseries 01} (2017) 121}
  [\href{https://arxiv.org/abs/1611.08426}{{\ttfamily 1611.08426}}].

\bibitem{Aalsma:2017ulu}
L.~Aalsma, J.~P. van~der Schaar and B.~Vercnocke, \emph{{Constrained
  superfields on metastable anti-D3-branes}},
  \href{https://doi.org/10.1007/JHEP05(2017)089}{\emph{JHEP} {\bfseries 05}
  (2017) 089} [\href{https://arxiv.org/abs/1703.05771}{{\ttfamily
  1703.05771}}].

\bibitem{Kallosh:2017wnt}
R.~Kallosh, A.~Linde, D.~Roest and Y.~Yamada, \emph{{$ \overline{D3} $ induced
  geometric inflation}},
  \href{https://doi.org/10.1007/JHEP07(2017)057}{\emph{JHEP} {\bfseries 07}
  (2017) 057} [\href{https://arxiv.org/abs/1705.09247}{{\ttfamily
  1705.09247}}].

\bibitem{GarciadelMoral:2017vnz}
M.~P. Garcia~del Moral, S.~Parameswaran, N.~Quiroz and I.~Zavala,
  \emph{{Anti-D3 branes and moduli in non-linear supergravity}},
  \href{https://doi.org/10.1007/JHEP10(2017)185}{\emph{JHEP} {\bfseries 10}
  (2017) 185} [\href{https://arxiv.org/abs/1707.07059}{{\ttfamily
  1707.07059}}].

\bibitem{Cribiori:2017laj}
N.~Cribiori, F.~Farakos, M.~Tournoy and A.~van Proeyen, \emph{{Fayet-Iliopoulos
  terms in supergravity without gauged R-symmetry}},
  \href{https://doi.org/10.1007/JHEP04(2018)032}{\emph{JHEP} {\bfseries 04}
  (2018) 032} [\href{https://arxiv.org/abs/1712.08601}{{\ttfamily
  1712.08601}}].

\bibitem{Kitazawa:2018zys}
N.~Kitazawa, \emph{{Brane SUSY Breaking and the Gravitino Mass}},
  \href{https://doi.org/10.1007/JHEP04(2018)081}{\emph{JHEP} {\bfseries 04}
  (2018) 081} [\href{https://arxiv.org/abs/1802.03088}{{\ttfamily
  1802.03088}}].

\bibitem{Krishnan:2018udc}
C.~Krishnan, H.~Raj and P.~N. Bala~Subramanian, \emph{{On the KKLT Goldstino}},
  \href{https://doi.org/10.1007/JHEP06(2018)092}{\emph{JHEP} {\bfseries 06}
  (2018) 092} [\href{https://arxiv.org/abs/1803.04905}{{\ttfamily
  1803.04905}}].

\bibitem{Aalsma:2018pll}
L.~Aalsma, M.~Tournoy, J.~P. Van Der~Schaar and B.~Vercnocke,
  \emph{{Supersymmetric embedding of antibrane polarization}},
  \href{https://doi.org/10.1103/PhysRevD.98.086019}{\emph{Phys. Rev.}
  {\bfseries D98} (2018) 086019}
  [\href{https://arxiv.org/abs/1807.03303}{{\ttfamily 1807.03303}}].

\bibitem{Cribiori:2018dlc}
N.~Cribiori, F.~Farakos and M.~Tournoy, \emph{{Supersymmetric Born-Infeld
  actions and new Fayet-Iliopoulos terms}},
  \href{https://doi.org/10.1007/JHEP03(2019)050}{\emph{JHEP} {\bfseries 03}
  (2019) 050} [\href{https://arxiv.org/abs/1811.08424}{{\ttfamily
  1811.08424}}].

\bibitem{Grana:2002tu}
M.~Grana, \emph{{D3-brane action in a supergravity background: The Fermionic
  story}}, \href{https://doi.org/10.1103/PhysRevD.66.045014}{\emph{Phys. Rev.}
  {\bfseries D66} (2002) 045014}
  [\href{https://arxiv.org/abs/hep-th/0202118}{{\ttfamily hep-th/0202118}}].

\bibitem{Grana:2003ek}
M.~Grana, T.~W. Grimm, H.~Jockers and J.~Louis, \emph{{Soft supersymmetry
  breaking in Calabi-Yau orientifolds with D-branes and fluxes}},
  \href{https://doi.org/10.1016/j.nuclphysb.2004.04.021}{\emph{Nucl. Phys.}
  {\bfseries B690} (2004) 21}
  [\href{https://arxiv.org/abs/hep-th/0312232}{{\ttfamily hep-th/0312232}}].

\bibitem{Marolf:2003ye}
D.~Marolf, L.~Martucci and P.~J. Silva, \emph{{Fermions, T duality and
  effective actions for D-branes in bosonic backgrounds}},
  \href{https://doi.org/10.1088/1126-6708/2003/04/051}{\emph{JHEP} {\bfseries
  04} (2003) 051} [\href{https://arxiv.org/abs/hep-th/0303209}{{\ttfamily
  hep-th/0303209}}].

\bibitem{Tripathy:2005hv}
P.~K. Tripathy and S.~P. Trivedi, \emph{{D3 brane action and fermion zero modes
  in presence of background flux}},
  \href{https://doi.org/10.1088/1126-6708/2005/06/066}{\emph{JHEP} {\bfseries
  06} (2005) 066} [\href{https://arxiv.org/abs/hep-th/0503072}{{\ttfamily
  hep-th/0503072}}].

\bibitem{Martucci:2005rb}
L.~Martucci, J.~Rosseel, D.~Van~den Bleeken and A.~Van~Proeyen, \emph{{Dirac
  actions for D-branes on backgrounds with fluxes}},
  \href{https://doi.org/10.1088/0264-9381/22/13/014}{\emph{Class. Quant. Grav.}
  {\bfseries 22} (2005) 2745}
  [\href{https://arxiv.org/abs/hep-th/0504041}{{\ttfamily hep-th/0504041}}].

\bibitem{Bergshoeff:2013pia}
E.~Bergshoeff, F.~Coomans, R.~Kallosh, C.~S. Shahbazi and A.~Van~Proeyen,
  \emph{{Dirac-Born-Infeld-Volkov-Akulov and Deformation of Supersymmetry}},
  \href{https://doi.org/10.1007/JHEP08(2013)100}{\emph{JHEP} {\bfseries 08}
  (2013) 100} [\href{https://arxiv.org/abs/1303.5662}{{\ttfamily 1303.5662}}].

\bibitem{Volkov:1973ix}
D.~V. Volkov and V.~P. Akulov, \emph{{Is the Neutrino a Goldstone Particle?}},
  \href{https://doi.org/10.1016/0370-2693(73)90490-5}{\emph{Phys. Lett.}
  {\bfseries B46} (1973) 109}.

\bibitem{Giddings:2001yu}
S.~B. Giddings, S.~Kachru and J.~Polchinski, \emph{{Hierarchies from fluxes in
  string compactifications}},
  \href{https://doi.org/10.1103/PhysRevD.66.106006}{\emph{Phys. Rev.}
  {\bfseries D66} (2002) 106006}
  [\href{https://arxiv.org/abs/hep-th/0105097}{{\ttfamily hep-th/0105097}}].

\bibitem{Polchinski:1998rr}
J.~Polchinski, \emph{{String theory. Vol. 2: Superstring theory and beyond}}.
  Cambridge University Press, 2007.

\bibitem{Frey:2008xw}
A.~R. Frey, G.~Torroba, B.~Underwood and M.~R. Douglas, \emph{{The Universal
  Kahler Modulus in Warped Compactifications}},
  \href{https://doi.org/10.1088/1126-6708/2009/01/036}{\emph{JHEP} {\bfseries
  01} (2009) 036} [\href{https://arxiv.org/abs/0810.5768}{{\ttfamily
  0810.5768}}].

\bibitem{Giddings:2005ff}
S.~B. Giddings and A.~Maharana, \emph{{Dynamics of warped compactifications and
  the shape of the warped landscape}},
  \href{https://doi.org/10.1103/PhysRevD.73.126003}{\emph{Phys. Rev.}
  {\bfseries D73} (2006) 126003}
  [\href{https://arxiv.org/abs/hep-th/0507158}{{\ttfamily hep-th/0507158}}].

\bibitem{deAlwis:2016cty}
S.~P. de~Alwis, \emph{{Constraints on Dbar Uplifts}},
  \href{https://doi.org/10.1007/JHEP11(2016)045}{\emph{JHEP} {\bfseries 11}
  (2016) 045} [\href{https://arxiv.org/abs/1605.06456}{{\ttfamily
  1605.06456}}].

\bibitem{Cownden:2016hpf}
B.~Cownden, A.~R. Frey, M.~C.~D. Marsh and B.~Underwood, \emph{{Dimensional
  Reduction for D3-brane Moduli}},
  \href{https://doi.org/10.1007/JHEP12(2016)139}{\emph{JHEP} {\bfseries 12}
  (2016) 139} [\href{https://arxiv.org/abs/1609.05904}{{\ttfamily
  1609.05904}}].

\bibitem{McGuirk:2012sb}
P.~McGuirk, G.~Shiu and F.~Ye, \emph{{Soft branes in supersymmetry-breaking
  backgrounds}}, \href{https://doi.org/10.1007/JHEP07(2012)188}{\emph{JHEP}
  {\bfseries 07} (2012) 188} [\href{https://arxiv.org/abs/1206.0754}{{\ttfamily
  1206.0754}}].

\bibitem{DeWolfe:2002nn}
O.~DeWolfe and S.~B. Giddings, \emph{{Scales and hierarchies in warped
  compactifications and brane worlds}},
  \href{https://doi.org/10.1103/PhysRevD.67.066008}{\emph{Phys. Rev.}
  {\bfseries D67} (2003) 066008}
  [\href{https://arxiv.org/abs/hep-th/0208123}{{\ttfamily hep-th/0208123}}].

\bibitem{Baumann:2007ah}
D.~Baumann, A.~Dymarsky, I.~R. Klebanov and L.~McAllister, \emph{{Towards an
  Explicit Model of D-brane Inflation}},
  \href{https://doi.org/10.1088/1475-7516/2008/01/024}{\emph{JCAP} {\bfseries
  0801} (2008) 024} [\href{https://arxiv.org/abs/0706.0360}{{\ttfamily
  0706.0360}}].

\bibitem{Kachru:2003sx}
S.~Kachru, R.~Kallosh, A.~D. Linde, J.~M. Maldacena, L.~P. McAllister and S.~P.
  Trivedi, \emph{{Towards inflation in string theory}},
  \href{https://doi.org/10.1088/1475-7516/2003/10/013}{\emph{JCAP} {\bfseries
  0310} (2003) 013} [\href{https://arxiv.org/abs/hep-th/0308055}{{\ttfamily
  hep-th/0308055}}].

\bibitem{Moritz:2017xto}
J.~Moritz, A.~Retolaza and A.~Westphal, \emph{{Toward de Sitter space from ten
  dimensions}}, \href{https://doi.org/10.1103/PhysRevD.97.046010}{\emph{Phys.
  Rev.} {\bfseries D97} (2018) 046010}
  [\href{https://arxiv.org/abs/1707.08678}{{\ttfamily 1707.08678}}].

\bibitem{Moritz:2018sui}
J.~Moritz and T.~Van~Riet, \emph{{Racing through the swampland: de Sitter
  uplift vs weak gravity}},
  \href{https://doi.org/10.1007/JHEP09(2018)099}{\emph{JHEP} {\bfseries 09}
  (2018) 099} [\href{https://arxiv.org/abs/1805.00944}{{\ttfamily
  1805.00944}}].

\bibitem{Kallosh:2018wme}
R.~Kallosh, A.~Linde, E.~McDonough and M.~Scalisi, \emph{{de Sitter Vacua with
  a Nilpotent Superfield}},
  \href{https://doi.org/10.1002/prop.201800068}{\emph{Fortsch. Phys.}
  {\bfseries 2018} (2018) 1800068}
  [\href{https://arxiv.org/abs/1808.09428}{{\ttfamily 1808.09428}}].

\bibitem{Moritz:2018ani}
J.~Moritz, A.~Retolaza and A.~Westphal, \emph{{On uplifts by warped
  anti‐D3‐branes}},
  \href{https://doi.org/10.1002/prop.201800098}{\emph{Fortsch. Phys.}
  {\bfseries 67} (2019) 1800098}
  [\href{https://arxiv.org/abs/1809.06618}{{\ttfamily 1809.06618}}].

\bibitem{Kallosh:2018psh}
R.~Kallosh, A.~Linde, E.~McDonough and M.~Scalisi, \emph{{4D models of de
  Sitter uplift}},
  \href{https://doi.org/10.1103/PhysRevD.99.046006}{\emph{Phys. Rev.}
  {\bfseries D99} (2019) 046006}
  [\href{https://arxiv.org/abs/1809.09018}{{\ttfamily 1809.09018}}].

\bibitem{Gautason:2018gln}
F.~F. Gautason, V.~Van~Hemelryck and T.~Van~Riet, \emph{{The Tension between
  10D Supergravity and dS Uplifts}},
  \href{https://doi.org/10.1002/prop.201800091}{\emph{Fortsch. Phys.}
  {\bfseries 67} (2019) 1800091}
  [\href{https://arxiv.org/abs/1810.08518}{{\ttfamily 1810.08518}}].

\bibitem{Hamada:2018qef}
Y.~Hamada, A.~Hebecker, G.~Shiu and P.~Soler, \emph{{On brane gaugino
  condensates in 10d}},
  \href{https://doi.org/10.1007/JHEP04(2019)008}{\emph{JHEP} {\bfseries 04}
  (2019) 008} [\href{https://arxiv.org/abs/1812.06097}{{\ttfamily
  1812.06097}}].

\bibitem{Kallosh:2019axr}
R.~Kallosh, A.~Linde, E.~McDonough and M.~Scalisi, \emph{{dS Vacua and the
  Swampland}}, \href{https://doi.org/10.1007/JHEP03(2019)134}{\emph{JHEP}
  {\bfseries 03} (2019) 134}
  [\href{https://arxiv.org/abs/1901.02022}{{\ttfamily 1901.02022}}].

\bibitem{Kallosh:2019oxv}
R.~Kallosh, \emph{{Gaugino Condensation and Geometry of the Perfect Square}},
  \href{https://doi.org/10.1103/PhysRevD.99.066003}{\emph{Phys. Rev.}
  {\bfseries D99} (2019) 066003}
  [\href{https://arxiv.org/abs/1901.02023}{{\ttfamily 1901.02023}}].

\bibitem{Hamada:2019ack}
Y.~Hamada, A.~Hebecker, G.~Shiu and P.~Soler, \emph{{Understanding KKLT from a
  10d perspective}},  \href{https://arxiv.org/abs/1902.01410}{{\ttfamily
  1902.01410}}.

\bibitem{Carta:2019rhx}
F.~Carta, J.~Moritz and A.~Westphal, \emph{{Gaugino condensation and small
  uplifts in KKLT}},  \href{https://arxiv.org/abs/1902.01412}{{\ttfamily
  1902.01412}}.

\bibitem{Gautason:2019jwq}
F.~F. Gautason, V.~Van~Hemelryck, T.~Van~Riet and G.~Venken, \emph{{A 10d view
  on the KKLT AdS vacuum and uplifting}},
  \href{https://arxiv.org/abs/1902.01415}{{\ttfamily 1902.01415}}.

\bibitem{Klebanov:2000hb}
I.~R. Klebanov and M.~J. Strassler, \emph{{Supergravity and a confining gauge
  theory: Duality cascades and chi SB resolution of naked singularities}},
  \href{https://doi.org/10.1088/1126-6708/2000/08/052}{\emph{JHEP} {\bfseries
  08} (2000) 052} [\href{https://arxiv.org/abs/hep-th/0007191}{{\ttfamily
  hep-th/0007191}}].

\bibitem{Aharony:2005ez}
O.~Aharony, Y.~E. Antebi and M.~Berkooz, \emph{{Open string moduli in KKLT
  compactifications}},
  \href{https://doi.org/10.1103/PhysRevD.72.106009}{\emph{Phys. Rev.}
  {\bfseries D72} (2005) 106009}
  [\href{https://arxiv.org/abs/hep-th/0508080}{{\ttfamily hep-th/0508080}}].

\bibitem{Gandhi:2011id}
S.~Gandhi, L.~McAllister and S.~Sjors, \emph{{A Toolkit for Perturbing Flux
  Compactifications}},
  \href{https://doi.org/10.1007/JHEP12(2011)053}{\emph{JHEP} {\bfseries 12}
  (2011) 053} [\href{https://arxiv.org/abs/1106.0002}{{\ttfamily 1106.0002}}].

\bibitem{Gukov:1999ya}
S.~Gukov, C.~Vafa and E.~Witten, \emph{{CFT's from Calabi-Yau four folds}},
  \href{https://doi.org/10.1016/S0550-3213(01)00289-9,
  10.1016/S0550-3213(00)00373-4}{\emph{Nucl. Phys.} {\bfseries B584} (2000) 69}
  [\href{https://arxiv.org/abs/hep-th/9906070}{{\ttfamily hep-th/9906070}}].

\bibitem{Baumann:2010sx}
D.~Baumann, A.~Dymarsky, S.~Kachru, I.~R. Klebanov and L.~McAllister,
  \emph{{D3-brane Potentials from Fluxes in AdS/CFT}},
  \href{https://doi.org/10.1007/JHEP06(2010)072}{\emph{JHEP} {\bfseries 06}
  (2010) 072} [\href{https://arxiv.org/abs/1001.5028}{{\ttfamily 1001.5028}}].

\bibitem{Dymarsky:2010mf}
A.~Dymarsky and L.~Martucci, \emph{{D-brane non-perturbative effects and
  geometric deformations}},
  \href{https://doi.org/10.1007/JHEP04(2011)061}{\emph{JHEP} {\bfseries 04}
  (2011) 061} [\href{https://arxiv.org/abs/1012.4018}{{\ttfamily 1012.4018}}].

\bibitem{Bergshoeff:2005yp}
E.~Bergshoeff, R.~Kallosh, A.-K. Kashani-Poor, D.~Sorokin and A.~Tomasiello,
  \emph{{An Index for the Dirac operator on D3 branes with background fluxes}},
  \href{https://doi.org/10.1088/1126-6708/2005/10/102}{\emph{JHEP} {\bfseries
  10} (2005) 102} [\href{https://arxiv.org/abs/hep-th/0507069}{{\ttfamily
  hep-th/0507069}}].

\bibitem{freedman2012supergravity}
D.~Freedman and A.~Van~Proeyen, \emph{Supergravity}. Cambridge University
  Press, 2012.

\bibitem{Wess:1992cp}
J.~Wess and J.~Bagger, \emph{{Supersymmetry and supergravity}}. 1992.

\bibitem{Rocek:1978nb}
M.~Rocek, \emph{{Linearizing the Volkov-Akulov Model}},
  \href{https://doi.org/10.1103/PhysRevLett.41.451}{\emph{Phys. Rev. Lett.}
  {\bfseries 41} (1978) 451}.

\bibitem{Lindstrom:1979kq}
U.~Lindstrom and M.~Rocek, \emph{{Constrained local superfields}},
  \href{https://doi.org/10.1103/PhysRevD.19.2300}{\emph{Phys. Rev.} {\bfseries
  D19} (1979) 2300}.

\bibitem{Casalbuoni:1988xh}
R.~Casalbuoni, S.~De~Curtis, D.~Dominici, F.~Feruglio and R.~Gatto,
  \emph{{Nonlinear Realization of Supersymmetry Algebra From Supersymmetric
  Constraint}}, \href{https://doi.org/10.1016/0370-2693(89)90788-0}{\emph{Phys.
  Lett.} {\bfseries B220} (1989) 569}.

\bibitem{Komargodski:2009rz}
Z.~Komargodski and N.~Seiberg, \emph{{From Linear SUSY to Constrained
  Superfields}},
  \href{https://doi.org/10.1088/1126-6708/2009/09/066}{\emph{JHEP} {\bfseries
  09} (2009) 066} [\href{https://arxiv.org/abs/0907.2441}{{\ttfamily
  0907.2441}}].

\bibitem{Kuzenko:2011tj}
S.~M. Kuzenko and S.~J. Tyler, \emph{{On the Goldstino actions and their
  symmetries}}, \href{https://doi.org/10.1007/JHEP05(2011)055}{\emph{JHEP}
  {\bfseries 05} (2011) 055} [\href{https://arxiv.org/abs/1102.3043}{{\ttfamily
  1102.3043}}].

\bibitem{Farakos:2013ih}
F.~Farakos and A.~Kehagias, \emph{{Decoupling Limits of sGoldstino Modes in
  Global and Local Supersymmetry}},
  \href{https://doi.org/10.1016/j.physletb.2013.06.001}{\emph{Phys. Lett.}
  {\bfseries B724} (2013) 322}
  [\href{https://arxiv.org/abs/1302.0866}{{\ttfamily 1302.0866}}].

\bibitem{Dudas:2015eha}
E.~Dudas, S.~Ferrara, A.~Kehagias and A.~Sagnotti, \emph{{Properties of
  Nilpotent Supergravity}},
  \href{https://doi.org/10.1007/JHEP09(2015)217}{\emph{JHEP} {\bfseries 09}
  (2015) 217} [\href{https://arxiv.org/abs/1507.07842}{{\ttfamily
  1507.07842}}].

\bibitem{Bergshoeff:2015tra}
E.~A. Bergshoeff, D.~Z. Freedman, R.~Kallosh and A.~Van~Proeyen, \emph{{Pure de
  Sitter Supergravity}}, \href{https://doi.org/10.1103/PhysRevD.93.069901,
  10.1103/PhysRevD.92.085040}{\emph{Phys. Rev.} {\bfseries D92} (2015) 085040}
  [\href{https://arxiv.org/abs/1507.08264}{{\ttfamily 1507.08264}}].

\bibitem{Hasegawa:2015bza}
F.~Hasegawa and Y.~Yamada, \emph{{Component action of nilpotent multiplet
  coupled to matter in 4 dimensional $ \mathcal{N}=1 $ supergravity}},
  \href{https://doi.org/10.1007/JHEP10(2015)106}{\emph{JHEP} {\bfseries 10}
  (2015) 106} [\href{https://arxiv.org/abs/1507.08619}{{\ttfamily
  1507.08619}}].

\bibitem{Ferrara:2015gta}
S.~Ferrara, M.~Porrati and A.~Sagnotti, \emph{{Scale invariant Volkov-Akulov
  supergravity}},
  \href{https://doi.org/10.1016/j.physletb.2015.08.066}{\emph{Phys. Lett.}
  {\bfseries B749} (2015) 589}
  [\href{https://arxiv.org/abs/1508.02939}{{\ttfamily 1508.02939}}].

\bibitem{DallAgata:2016yof}
G.~Dall'Agata, E.~Dudas and F.~Farakos, \emph{{On the origin of constrained
  superfields}}, \href{https://doi.org/10.1007/JHEP05(2016)041}{\emph{JHEP}
  {\bfseries 05} (2016) 041}
  [\href{https://arxiv.org/abs/1603.03416}{{\ttfamily 1603.03416}}].

\bibitem{Ferrara:2016een}
S.~Ferrara, R.~Kallosh, A.~Van~Proeyen and T.~Wrase, \emph{{Linear Versus
  Non-linear Supersymmetry, in General}},
  \href{https://doi.org/10.1007/JHEP04(2016)065}{\emph{JHEP} {\bfseries 04}
  (2016) 065} [\href{https://arxiv.org/abs/1603.02653}{{\ttfamily
  1603.02653}}].

\bibitem{Brignole:1997pe}
A.~Brignole, F.~Feruglio and F.~Zwirner, \emph{{On the effective interactions
  of a light gravitino with matter fermions}},
  \href{https://doi.org/10.1088/1126-6708/1997/11/001}{\emph{JHEP} {\bfseries
  11} (1997) 001} [\href{https://arxiv.org/abs/hep-th/9709111}{{\ttfamily
  hep-th/9709111}}].

\bibitem{DallAgata:2015pdd}
G.~Dall'Agata, S.~Ferrara and F.~Zwirner, \emph{{Minimal scalar-less
  matter-coupled supergravity}},
  \href{https://doi.org/10.1016/j.physletb.2015.11.066}{\emph{Phys. Lett.}
  {\bfseries B752} (2016) 263}
  [\href{https://arxiv.org/abs/1509.06345}{{\ttfamily 1509.06345}}].

\bibitem{Kuzenko:2018jlz}
S.~M. Kuzenko, \emph{{Taking a vector supermultiplet apart: Alternative
  Fayet-Iliopoulos-type terms}},
  \href{https://doi.org/10.1016/j.physletb.2018.04.051}{\emph{Phys. Lett.}
  {\bfseries B781} (2018) 723}
  [\href{https://arxiv.org/abs/1801.04794}{{\ttfamily 1801.04794}}].

\bibitem{Antoniadis:2018cpq}
I.~Antoniadis, A.~Chatrabhuti, H.~Isono and R.~Knoops, \emph{{Fayet-Iliopoulos
  terms in supergravity and D-term inflation}},
  \href{https://doi.org/10.1140/epjc/s10052-018-5861-6}{\emph{Eur. Phys. J.}
  {\bfseries C78} (2018) 366}
  [\href{https://arxiv.org/abs/1803.03817}{{\ttfamily 1803.03817}}].

\bibitem{Antoniadis:2018oeh}
I.~Antoniadis, A.~Chatrabhuti, H.~Isono and R.~Knoops, \emph{{The cosmological
  constant in Supergravity}},
  \href{https://doi.org/10.1140/epjc/s10052-018-6175-4}{\emph{Eur. Phys. J.}
  {\bfseries C78} (2018) 718}
  [\href{https://arxiv.org/abs/1805.00852}{{\ttfamily 1805.00852}}].

\bibitem{Aldabergenov:2018nzd}
Y.~Aldabergenov, S.~V. Ketov and R.~Knoops, \emph{{General couplings of a
  vector multiplet in $N=1$ supergravity with new FI terms}},
  \href{https://doi.org/10.1016/j.physletb.2018.07.072}{\emph{Phys. Lett.}
  {\bfseries B785} (2018) 284}
  [\href{https://arxiv.org/abs/1806.04290}{{\ttfamily 1806.04290}}].

\bibitem{Aldabergenov:2019hvl}
Y.~Aldabergenov, \emph{{No-scale supergravity with new Fayet-Iliopoulos term}},
   \href{https://arxiv.org/abs/1903.11829}{{\ttfamily 1903.11829}}.

\bibitem{Kuzenko:2019vaw}
S.~M. Kuzenko, \emph{{Superconformal vector multiplet self-couplings and
  generalised Fayet-Iliopoulos terms}},
  \href{https://arxiv.org/abs/1904.05201}{{\ttfamily 1904.05201}}.

\bibitem{Gaillard:1981rj}
M.~K. Gaillard and B.~Zumino, \emph{{Duality Rotations for Interacting
  Fields}}, \href{https://doi.org/10.1016/0550-3213(81)90527-7}{\emph{Nucl.
  Phys.} {\bfseries B193} (1981) 221}.

\bibitem{Kugo:1983mv}
T.~Kugo and S.~Uehara, \emph{{$N=1$ Superconformal Tensor Calculus: Multiplets
  With External Lorentz Indices and Spinor Derivative Operators}},
  \href{https://doi.org/10.1143/PTP.73.235}{\emph{Prog. Theor. Phys.}
  {\bfseries 73} (1985) 235}.

\bibitem{Dall'Agata:2015lek}
G.~Dall'Agata and F.~Farakos, \emph{{Constrained superfields in Supergravity}},
  \href{https://doi.org/10.1007/JHEP02(2016)101}{\emph{JHEP} {\bfseries 02}
  (2016) 101} [\href{https://arxiv.org/abs/1512.02158}{{\ttfamily
  1512.02158}}].

\bibitem{Schillo:2015ssx}
M.~Schillo, E.~van~der Woerd and T.~Wrase, \emph{{The general de Sitter
  supergravity component action}}, \href{https://doi.org/10.1002/prop201500074,
  10.1002/prop.201500074}{\emph{Fortsch. Phys.} {\bfseries 64} (2016) 292}
  [\href{https://arxiv.org/abs/1511.01542}{{\ttfamily 1511.01542}}].

\bibitem{Yamada:2016tca}
Y.~Yamada, \emph{{Construction of higher-derivative supergravity models via
  superconformal formulation}}, Ph.D. thesis, Waseda U., 2016-02.

\bibitem{Cribiori:2017ngp}
N.~Cribiori, G.~Dall'Agata and F.~Farakos, \emph{{From Linear to Non-linear
  SUSY and Back Again}},
  \href{https://doi.org/10.1007/JHEP08(2017)117}{\emph{JHEP} {\bfseries 08}
  (2017) 117} [\href{https://arxiv.org/abs/1704.07387}{{\ttfamily
  1704.07387}}].

\bibitem{Cribiori:2019cgz}
N.~Cribiori, \emph{{Non-linear realisations in global and local
  supersymmetry}}, Ph.D. thesis, Padua U., Astron. Dept., 2018.
\newblock \href{https://arxiv.org/abs/1901.02097}{{\ttfamily 1901.02097}}.

\bibitem{Aparicio:2015psl}
L.~Aparicio, F.~Quevedo and R.~Valandro, \emph{{Moduli Stabilisation with
  Nilpotent Goldstino: Vacuum Structure and SUSY Breaking}},
  \href{https://doi.org/10.1007/JHEP03(2016)036}{\emph{JHEP} {\bfseries 03}
  (2016) 036} [\href{https://arxiv.org/abs/1511.08105}{{\ttfamily
  1511.08105}}].

\bibitem{Balasubramanian:2005zx}
V.~Balasubramanian, P.~Berglund, J.~P. Conlon and F.~Quevedo,
  \emph{{Systematics of moduli stabilisation in Calabi-Yau flux
  compactifications}},
  \href{https://doi.org/10.1088/1126-6708/2005/03/007}{\emph{JHEP} {\bfseries
  03} (2005) 007} [\href{https://arxiv.org/abs/hep-th/0502058}{{\ttfamily
  hep-th/0502058}}].

\bibitem{Conlon:2005ki}
J.~P. Conlon, F.~Quevedo and K.~Suruliz, \emph{{Large-volume flux
  compactifications: Moduli spectrum and D3/D7 soft supersymmetry breaking}},
  \href{https://doi.org/10.1088/1126-6708/2005/08/007}{\emph{JHEP} {\bfseries
  08} (2005) 007} [\href{https://arxiv.org/abs/hep-th/0505076}{{\ttfamily
  hep-th/0505076}}].

\bibitem{Kallosh:2018nrk}
R.~Kallosh and T.~Wrase, \emph{{dS Supergravity from 10d}},
  \href{https://doi.org/10.1002/prop.201800071}{\emph{Fortsch. Phys.}
  {\bfseries 2018} (2018) 1800071}
  [\href{https://arxiv.org/abs/1808.09427}{{\ttfamily 1808.09427}}].

\bibitem{Jockers:2004yj}
H.~Jockers and J.~Louis, \emph{{The Effective action of D7-branes in N = 1
  Calabi-Yau orientifolds}},
  \href{https://doi.org/10.1016/j.nuclphysb.2004.11.009}{\emph{Nucl. Phys.}
  {\bfseries B705} (2005) 167}
  [\href{https://arxiv.org/abs/hep-th/0409098}{{\ttfamily hep-th/0409098}}].

\bibitem{Grimm:2011dx}
T.~W. Grimm and D.~Vieira~Lopes, \emph{{The N=1 effective actions of D-branes
  in Type IIA and IIB orientifolds}},
  \href{https://doi.org/10.1016/j.nuclphysb.2011.10.019}{\emph{Nucl. Phys.}
  {\bfseries B855} (2012) 639}
  [\href{https://arxiv.org/abs/1104.2328}{{\ttfamily 1104.2328}}].

\bibitem{Kerstan:2011dy}
M.~Kerstan and T.~Weigand, \emph{{The Effective action of D6-branes in N=1 type
  IIA orientifolds}},
  \href{https://doi.org/10.1007/JHEP06(2011)105}{\emph{JHEP} {\bfseries 06}
  (2011) 105} [\href{https://arxiv.org/abs/1104.2329}{{\ttfamily 1104.2329}}].

\bibitem{Carta:2016ynn}
F.~Carta, F.~Marchesano, W.~Staessens and G.~Zoccarato, \emph{{Open string
  multi-branched and K\"ahler potentials}},
  \href{https://doi.org/10.1007/JHEP09(2016)062}{\emph{JHEP} {\bfseries 09}
  (2016) 062} [\href{https://arxiv.org/abs/1606.00508}{{\ttfamily
  1606.00508}}].

\bibitem{Escobar:2018tiu}
D.~Escobar, F.~Marchesano and W.~Staessens, \emph{{Type IIA Flux Vacua with
  Mobile D6-branes}},
  \href{https://doi.org/10.1007/JHEP01(2019)096}{\emph{JHEP} {\bfseries 01}
  (2019) 096} [\href{https://arxiv.org/abs/1811.09282}{{\ttfamily
  1811.09282}}].

\bibitem{Cribiori:2016hdz}
N.~Cribiori, G.~Dall'Agata and F.~Farakos, \emph{{Interactions of N Goldstini
  in Superspace}},
  \href{https://doi.org/10.1103/PhysRevD.94.065019}{\emph{Phys. Rev.}
  {\bfseries D94} (2016) 065019}
  [\href{https://arxiv.org/abs/1607.01277}{{\ttfamily 1607.01277}}].

\bibitem{Samuel:1982uh}
S.~Samuel and J.~Wess, \emph{{A Superfield Formulation of the Nonlinear
  Realization of Supersymmetry and Its Coupling to Supergravity}},
  \href{https://doi.org/10.1016/0550-3213(83)90622-3}{\emph{Nucl. Phys.}
  {\bfseries B221} (1983) 153}.

\bibitem{Farakos:2018sgq}
F.~Farakos, A.~Kehagias and A.~Riotto, \emph{{Liberated $ \mathcal{N} $ = 1
  supergravity}}, \href{https://doi.org/10.1007/JHEP06(2018)011}{\emph{JHEP}
  {\bfseries 06} (2018) 011}
  [\href{https://arxiv.org/abs/1805.01877}{{\ttfamily 1805.01877}}].

\end{thebibliography}

\providecommand{\href}[2]{#2}\begingroup\raggedright\endgroup

\end{document}